\documentclass[%
reprint, superscriptaddress,
showpacs,
 amsmath,amssymb,
 aps,
]{revtex4-1}

\usepackage[nodisplayskipstretch]{setspace}
\usepackage{cancel}
\usepackage{mathtools}
\allowdisplaybreaks
\usepackage{tabularx}

\usepackage{amssymb}
\usepackage{amsmath}
\usepackage{epsfig}
\usepackage[hidelinks]{hyperref}
\usepackage{breakurl}
\usepackage{bm}
\usepackage{color}
\usepackage{tikz}

\usepackage[caption=false]{subfig}
\usepackage[english]{babel}
\usepackage{natbib}

\newcommand{\F}{\mathcal{F}}
\newcommand{\G}{\mathcal{G}}

\DeclareMathAlphabet\mathbfcal{OMS}{cmsy}{b}{n}
\allowdisplaybreaks

\begin{document}
\title{
Third order post-Newtonian gravitational radiation from two-body scattering. Instantaneous Energy and Angular momentum radiations
}
\date{\today}
\author{Gihyuk Cho}
\email{gihyuk.cho@desy.de}
\affiliation{Deutsches Elektronen-Synchrotron DESY, Notkestr. 85, 22607 Hamburg, Germany}

\author{Subhajit Dandapat}
\email{subhajit.dandapat@tifr.res.in}
\affiliation{Department of Astronomy and Astrophysics,
 Tata Institute of Fundamental Research, Mumbai 400005, India}
 
\author{Achamveedu Gopakumar}
\affiliation{Department of Astronomy and Astrophysics,
 Tata Institute of Fundamental Research, Mumbai 400005, India}
 
\begin{abstract}
We compute the third post-Newtonian (3PN) accurate instantaneous contributions to the radiated gravitational wave (GW) energy and angular momentum arising 
from the hyperbolic passages of non-spinning compact objects.
The present computations employ 3PN-accurate instantaneous contributions to the far-zone energy and angular momentum fluxes and the 3PN-accurate 
Keplerian type parametric solution for compact binaries in hyperbolic orbits. 
\end{abstract}

\maketitle
\section{introduction}
The routine detection of 
transient GW events that arise from merging black hole (BH) binaries in bound orbits has inaugurated the era of GW astronomy \cite{abbott2019gwtc,abbott2021gwtc,venumadhav2020new}. 
Further, observations of a neutron star binary coalescence in GWs, and many electromagnetic frequency windows have 
provided a peak into  the benefits of the multi-messenger GW astronomy \cite{abbott2017gw170817,Poggiani:2019enk,monitor2017gravitational}.
\ In contrast,
compact binaries in unbound orbits can provide transient GW burst events in the LIGO,LISA and IPTA GW frequency windows \cite{garcia2018gravitational,mukherjee2020detectability,kocsis2006detection,burke2019astrophysics}. 
Interestingly,  GW burst events due to hyperbolic encounters of neutron stars may even be accompanied by electromagnetic
flares \cite{tsang2013shattering}.
\ Therefore, there are on-going PN efforts to characterize 
both the dynamics and associated GW emission aspects of compact binaries in hyperbolic orbits in general relativity \cite{cho2018gravitational,bae2020gravitational,bini2021radiative,bini2021gravitational,bini2021higher,bini2021frequency}.
\par
The present effort extends the classic computations of the radiated 
energy($\Delta {\cal E} $) and angular momentum 
 ($\Delta {\cal J}$) during hyperbolic encounters 
 of non-spinning compact objects 
 to 1PN order \cite{Hansen:1972jt,blanchet1989higher,junker1992binary}. 
 Recall that PN approximation allows us to write, for example, the orbital dynamics of 
 non-spinning compact binaries 
as corrections to Newtonian equations of motion
in powers of $ ( v/c)^2 \sim G\, M/(c^2\,r)$, where $v, M,$ and $r$ are
the velocity, total mass and relative separation of the binary \cite{blanchet2014gravitational,Porto:2016pyg} with the gravitation constant $G$ and speed of light $c$. 
Note that the expressions, available in Refs.\cite{blanchet1989higher,junker1992binary},
provided the next-to-leading order (1PN) contributions to $\Delta {\cal E} $ and $\Delta {\cal J}$, influenced by Refs.~\cite{1971ESRSP..52...45R,wagoner1976post}.
The present computation provides 
3PN-accurate `instantaneous' contributions to 
$\Delta {\cal E} $ and $\Delta {\cal J}$  with the help of 
Refs.~\cite{arun2008inspiralling,arun2009third,cho2018gravitational}
in the modified Harmonic gauge.
It turned out that at  PN orders beyond the 1PN,  the radiative moments and the resulting far-zone fluxes  have two distinct contributions \cite{blanchet2014gravitational}. 
One part of the radiative moments and their fluxes depends only at the usual retarded time and it is customary to  refer these terms  as the
“instantaneous contributions”. In contrast, 
the second part depends on the dynamics of compact binary in its entire past and therefore these contributions are 
usually termed as the “hereditary contributions”\cite{Blanchet:1987wq,Blanchet:1992br}.
In this paper, we focus our efforts on the instantaneous contributions.\\
The present computations are not a straightforward 
extensions to 3PN order  of what are done in Refs.~\cite{blanchet1989higher,junker1992binary}.
This is mainly because of logarithmic terms that appear at the 3PN corrections to far-zone energy and angular momentum fluxes associated with compact binaries in non-circular orbits  \cite{arun2008inspiralling,arun2009third}. 
We provide a prescription to compute these logarithmic integrals in terms of Clausen function of order two \cite{lewin1991structural}.
%
\par
 The  manuscript is structured in the following way. 
 The way of our PN-accurate computations and underlying formalism to obtain
 $\Delta {\cal E} $ and $\Delta {\cal J}$ and the results are presented in Sec.~\ref{main}. And also we briefly present parabolic limit and the implications of the bremsstrahlung limit.
The detail of our computations are presented in the 
Appendix.~ \ref{log_integral}, and our PN-accurate expressions in terms of energy and angular momentum is given in Appendix.~\ref{appendix_AC2}
while Sec.~\ref{summary} provides a brief summary and  on-going investigations.
 \section{3PN accurate Instantaneous Contributions 
 to $\Delta {\cal E} $ and $\Delta {\cal J}$  }\label{main}
By the matching between multipolar post-Minkowskian (MPM) expansion and PN expansion \cite{Blanchet:1998in,
Poujade:2001ie}, gravitational fluxes $\cal{F}$ (either energy or angular momentum) can be expressed in terms of mechanical variables describing binaries such as mass, radial distance and velocities, of which explicit expression can be found in Eq.(5.2) in \cite{arun2008inspiralling} in standard/modified harmonic gauge. Additionally, once the 3PN accurate quasi-Keplerian solution \textit{i.e.} the mechanical variables in time (implicitly), obtained as in \cite{cho2018gravitational}, the fluxes can be fully written as a function of time $t$. Here, we briefly show the PN structure of the fluxes upto 3PN($1/c^6$) order,
\begin{subequations}
\begin{align}
\mathcal{F}(t)=\mathcal{F}_{\rm{inst}}(t)+\mathcal{F}_{\rm{hered}}(t)\,,
\end{align}
where
\begin{align}
&\mathcal{F}_{\rm{inst}}=\mathcal{F}_\text{N} +\frac{1}{c^2}\mathcal{F}_\text{1PN}+\frac{1}{c^4}\mathcal{F}_\text{1PN}+\frac{1}{c^6}\mathcal{F}_\text{1PN}+\mathcal{O}(1/c^7)\,,\\[1ex]
&\mathcal{F}_{\rm{hered}}=\frac{1}{c^3}\,\mathcal{F}_\text{tail} +\frac{1}{c^6}\,\mathcal{F}_\text{tail(tail)}+\frac{1}{c^6}\mathcal{F}_{\text{tail}^2}+\mathcal{O}(1/c^7)\,.
\end{align}
\end{subequations}
By $\mathcal{F}_{\rm{hered}}$ (hereditary contribution), we mean all contribution that is dependent on the past history of binaries \cite{Blanchet:1987wq,Blanchet:1992br}. Otherwise, it is called instantaneous contribution denoted as $\mathcal{F}_{\rm{inst}}$. Thus, total radiations $\int^{+\infty}_{-\infty} dt \,\mathcal{F}$ also have both instantaneous and hereditary contributions and the same PN structure. Note that total radiations are observables, hence gauge invariant but each instantaneous/hereditary contribution is not gauge invariant because of an ambiguous separation of the long and short scales leaving coordinate dependence via $r_0$ (which will be seen shortly). The full result should not be dependent on the scale $r_0$, and hence recover gauge invariance. The leading (Newtonian) order of energy and angular momentum radiations $\Delta\cal{E}$, $\Delta\cal{J}$ (hence leading order of instantaneous part because hereditary part starts at 1.5PN order), were computed in Ref.~\cite{Hansen:1972jt} for the first time. Its extension upto 1PN, (hence still instantaneous) was made in \cite{blanchet1989higher,junker1992binary}. The higher PN orders including hereditary contribution have never been treated so far. As one of serial works in the line of completing 3PN accurate radiations, we compute and complete the instantaneous contribution first as what follows.\\
In Sec.\ref{E_quad}, we will explain how the computation goes by an example at leading order influenced by Ref.~\cite{Hansen:1972jt,blanchet1989higher}. All computation is similar to the leading order one except the logarithmic terms that appear at the 3PN order. The detailed way of tackling these logarithmic terms is provided in Appendix \ref{log_integral}.
These computations are repeated to obtain angular momentum radiation
in Sec.~\ref{am_3pn}.
Thereafter, we explain 
why our instantaneous 
contributions to the radiated energy and angular momentum during hyperbolic encounters are exact up to 3PN order and explore 
their limiting cases.
\subsection{Newtonian order $\Delta {\cal E} $ and $\Delta {\cal J}$ Computations}
\label{E_quad}

We begin by explaining the procedure of Ref.~\cite{Hansen:1972jt,blanchet1989higher} for computing
the radiated energy and angular momentum in GWs during hyperbolic encounters at the Newtonian order.
The natural starting point is the familiar Newtonian(leading) order far-zone 
GW energy flux $\mathcal{F}_\text{N}$ for compact binaries in generic orbits \cite{blanchet1989higher}

\begin{align}
\mathcal{F}_\text{N}&=\frac{8}{15} \frac{G^3 M^2 \mu^2}{c^5 r^4} \left(12 \, v^2-11 \dot{r}^2\right)\,.
\end{align}
The approach of Ref.~\cite{blanchet1989higher} requires us to employ
the standard Keplerian parametric solution for compact binaries
in Newtonian hyperbolic orbits, available in Refs. ~\cite{AIHPA_1985__43_1_107_0,klioner2016basic}.
 This is for expressing the total orbital velocity $v$, the radial velocity $\dot r$ 
and the radial separation 
in terms 
of various elements of the Keplerian parametric solution for hyperbolic orbits.
The underlying parametric solution for hyperbolic orbits reads 
\begin{subequations}
\label{newtonian_hyper}
\begin{align}
r&=a \, (e \, \cosh u-1) \\
\phi-\phi_0&=2 \, \tan^{-1} \left( \sqrt{\frac{e+1}{e-1}} \, \tanh \frac{u}{2} \right)\\
n\,(t-t_0)&=e \, \sinh u -u\,,
 \end{align}
\end{subequations}
where $r$, $\phi$ and $t$ are radial orbital separation, the angular variable of the reduced mass $\mu$ around the total mass $M$ and coordinate time, respectively. 
Further, the eccentric anomaly 
 parameter $u$ has the range $-\infty<u<\infty$ while $\phi_0$ and $t_0$ denote some initial value of $\phi$ and $t$.
The familiar Newtonian semi-major axis $a$, orbital eccentricity $e$ and 
the mean motion $n$ are given in terms of the conserved reduced energy $E=\frac{\mathcal{E}}{\mu}$ and reduced angular momentum $j=\frac{\cal J}{\mu \, G \, M}$,
\begin{align*}
a&=\frac{G M}{2 E} \\
e^2&=1+2 \, E\,j^2 \\
n&=\frac{(2E)^{3/2}}{G M}\,.
\end{align*}
It is fairly straightforward to express 
 $\dot{r}^2$, $v^2$  in terms of $E, j^2, (e \cosh u-1)$ with the help of 
 \begin{subequations}
\begin{align}
\dot{r}^2&=\left(\frac{dr}{du} \bigg{/} \frac{dt}{du}\right)^2 \\
\dot{\phi}^2&=\left(\frac{d\phi}{du} \bigg{/}\frac{dt}{du}\right)^2 \\
v^2&=\dot{r}^2+r^2 \, \dot{\phi}^2.
\end{align}
\end{subequations}
This leads to 
\begin{subequations}
	\label{rvn}
\begin{align}
\dot{r}^2&=\bigg{(}1+ \frac{2}{(e \cosh u -1)}- \frac{(2Eh^2)}{(e \cosh u -1)^2}\bigg{)} (2 E)\,,\\[1ex]
v^2&=\left(1+ \frac{2}{(e \cosh u-1)}\right) (2 E).
\end{align}
\end{subequations}
Using Eqs.~\eqref{rvn},  we can express the instantaneous energy flux as a polynomial in $(1-e \, \cosh u)^{-1}$ and the final expression reads 

\begin{align}
\mathcal{F}_\text{N}&=\frac{1}{n}\,\frac{du}{dt}\sum_{N=3}^{5}  \frac{\bar{\alpha}_N(E,j)}{(e \cosh u-1)^{N}}\,,\\[1ex]
\bar{\alpha}_N (e_t)&=\frac{\nu^2}{G} \, \left(\frac{E}{c}\right)^5 \alpha_N (E,j)\,.
\end{align}
In the above equation $\nu$ is defined as dimensionless mass parameter of the binary, namely $\nu={m_1 \, m_2}/{(m_1+m_2)^2}$, where $m_1$ and $m_2$ are the masses of the binary configuration.
The coefficients $\alpha_N (E,j)$ at the Newtonian order are given by
\begin{align}
	\alpha_3&=\frac{256}{15} \nonumber\\[0.5ex] \alpha_4&=\frac{512}{15} \nonumber \\[0.5ex]
	\alpha_5&=\frac{5632}{15} \, E\, j^2 .
\end{align}
The fact that we have parametrized the far-zone Newtonian energy flux in terms of 
Newtonian hyperbolic orbital description allows us to write 
the total radiated energy in GWs at the Newtonian order as 
	

\begin{align}
\label{lh}
\Delta \mathcal{E}_\text{N}&=\int_{-\infty}^\infty \mathcal{F}_\text{N} \, dt\,, \nonumber \\[0.5ex]
&=\frac{1}{n} \sum_{N=3}^{5} \bar{\alpha}_N(E,j) \, \int_{-\infty}^\infty\frac{du}{(e_t \cosh u-1)^{N}}\,.
\end{align}

Clearly, we can easily obtain the desired expression for the quadrupolar order 
$\Delta \mathcal{E}$ during hyperbolic encounters if we can compute the three integrals 
that appear on the right-hand side of Eq.~\eqref{lh}.
With the help of Refs.~\cite{Hansen:1972jt,blanchet1989higher}, we 
find 
\begin{align}
\label{int}
&\int_{-\infty}^\infty\frac{du}{(e \cosh u-x)^{N}}\\[0.5ex]
&=\frac{2}{(N-1)!} \bigg{[}\bigg{(}\frac{d}{d x}\bigg{)}^{N-1} \bigg{(}\frac{1}{\sqrt{e^2-x^2}}\, \arccos \bigg{(}-\frac{x}{e}\bigg{)}\bigg{)}\bigg{]}\,.\notag
\end{align}
This leads to
\begin{subequations}
   
\label{enag_nw}
\begin{align}
\label{E_q}
\Delta \mathcal{E}_\text{N}&=
\frac{2 M \, \nu ^2}{15 c^5 \, j^7} 
\, \bigg{[}\sqrt{e^2-1} \bigg{(}\frac{602}{3}+\frac{673 \,  e^2}{3}\bigg{)} \nonumber \\
&+\bigg{(}96+292 \, e^2+37 \, e^4\bigg{)} \arccos \bigg{(}-\frac{1}{e}\bigg{)} \bigg{]}\,,
\end{align}
where we have used the following Newtonian accurate relation that connects 
$E$ to $j$ and $e$, namely $E=\frac{e^2-1}{2 \, j^2}$, to obtain the above result.
Indeed, our expression is fully consistent with Ref.~\cite{Hansen:1972jt,blanchet1989higher,junker1992binary}.


\end{subequations}

 We now move onto explain briefly how Ref.~\cite{junker1992binary} computed the 
  quadrupolar order contributions to the radiated angular momentum during 
  hyperbolic encounters of non-spinning compact objects.
We begin by the Newtonian order angular momentum flux $\mathcal{G}_\text{N}$ for compact binaries 
in generic orbits \cite{blanchet1989higher}:
\begin{align}
\label{ag}
\mathbfcal{G}_\text{N}&=\frac{8}{5} \,\mathbf{\tilde{L}_N} \,  \frac{G^2 M \mu^2}{c^5 \, r^3} \left\{2v^2-3 \dot{r}^2+2 \, \frac{G M}{r}\right\} 
\end{align}
where $\mathbf{\tilde{L}_N}=\mathbf{r \times v}$ stands for the scaled Newtonian 
angular momentum vector.
The fact that the 
the orbital angular momentum vector remains a constant as we consider only non-spinning compact binaries allows to us to compute 
an expression for the  angular momentum flux $d {\cal J}/dt$ from the above equation.
Thereafter, we pursue the steps involved in the 
$\Delta {\cal E}$ computations and 
this leads to
\begin{align}
\label{G_N}
\mathcal{G}_\text{N}&=\frac{1}{n}\frac{du}{dt}\sum_{N=2}^{4}  \frac{\bar{A}_N(E,j)}{(e \cosh u-1)^{N}}\,,\\[1ex]
\bar{A}_N (e_t)&= \frac{M \, j \, E^4 \, \nu^2}{c^5} \, A_N (E,j) \,.
\end{align}
The three constant coefficients are given by 
\begin{align}
	A_2&=-\frac{128}{5}\,,\notag\\[0.5ex]
	A_3&=0\,, \nonumber \\[0.5ex]
	A_4&=\frac{768}{5} \, E\, j^2 \,.
\end{align}
The radiated angular momentum $\Delta \mathcal{J}_\text{N}$ at the Newtonian order becomes
\begin{align}
\label{jh}
\Delta \mathcal{J}_\text{N}&=\int_{-\infty}^\infty \mathcal{G}_\text{N} \, dt \nonumber\,, \\[1ex]
&=\frac{1}{n} \sum_{N=2}^{4} \bar{A}_N(E,j) \, \int_{-\infty}^\infty\frac{du}{(e_t \cosh u-1)^{N}}.
\end{align}
Clearly, these integrals are similar to those we tackled earlier and 
this eventually leads to 
\begin{align}
\Delta \mathcal{J}_\text{N}&=
\frac{8 \, G M^2 \nu ^2}{5 \, c^5 \, j^4}  \, 
 \,  \bigg{\{}(13+2 \, e^2) \,  \sqrt{e^2-1}\nonumber \\[0.5ex]
 &+(8+7 \, e^2) \arccos \left( -\frac{1}{e}\right)\bigg{\}}.
\end{align}
We have verified that the above expression is fully consistent with Ref.~\cite{Hansen:1972jt,blanchet1989higher,junker1992binary}.
We now move on to extend these calculations to 3PN order while focusing on the 
instantaneous contributions.

\subsection{ 3PN-accurate instantaneous contributions to the radiated energy}
\label{inst_en}

 It should be obvious that we require two crucial ingredients for our 3PN-accurate  $\Delta {\cal E}$ computation.
 The first ingredient is the 3PN accurate `instantaneous' contributions to 
 the far-zone GW energy flux from non-spinning compact binaries 
 in non-circular orbits \cite{blanchet2014gravitational}. These instantaneous contributions depend only on the state of the binary at
the usual retarded time and they appear usually at the Newtonian, 1PN, 2PN, 2.5PN and 3PN orders. 
In contrast,
the hereditary contributions, as the name suggests, are sensitive to the binary dynamics at all epochs prior to the usual retarded time 
and appear at 1.5PN (relative) order for the first time \cite{blanchet2014gravitational}. In this paper, we focus our efforts on 
the 3PN-accurate `instantaneous' contributions to the far-zone fluxes, as given 
by Eqs.~(5.2) in  Ref.~\cite{arun2008inspiralling} 
in the Modified harmonic (MH) coordinates.
A close inspection of these contributions reveal that $\Delta {\cal E}$ computation
at 3PN order will be demanding due to the presence of certain `logarithmic' terms, as evident from Eq.(5.2e) of Ref.~\cite{arun2008inspiralling}. 
The second ingredient for the present computation is the 3PN-accurate 
generalized quasi-Keplerian parametric solution for compact binaries 
in hyperbolic orbits, derived in Ref.~\cite{cho2018gravitational}.


Note that Ref.~\cite{cho2018gravitational} provided a parametric way to 
track 3PN-accurate conservative trajectory of compact binaries in hyperbolic orbits.
This effort extended the 1PN-accurate derivation of Keplerian type parametric 
solution for hyperbolic motion that employed the arguments of analytic continuation \cite{AIHPA_1985__43_1_107_0}.
At the 3PN order, the radial motion $r(t)$ is conveniently parametrized as 
		\begin{subequations}
	\begin{align}
	    &r=a_r (e_r \cosh u -1) \\
	    &n(t-t_0)=e_t \sinh u -u+\left( \frac{f_{4t}}{c^4}+\frac{f_{6t}}{c^6} \right) V\notag \\[1ex]
	    &+\left( \frac{g_{4t}}{c^4}+\frac{g_{6t}}{c^6} \right) \sin V+\frac{h_{6t}}{c^6} \sin 2V+\frac{i_{6t}}{c^6} \sin 3V,
	\end{align}
\end{subequations}
where $u$ is the eccentric anomaly while  $a_r,\ e_r,\ e_t,\ n $ and $t_0$ are certain PN-accurate semi-major axis, radial eccentricity, time eccentricity, mean motion, and initial epoch, respectively. In addition, we have several orbital functions like 
$g_{4t},g_{6t},f_{4t},f_{6t},i_{6t}$ and $h_{6t}$ that appear at 2PN and 3PN orders.
Further, the angular motion is described by  
\begin{subequations}
\begin{align}
    &\phi-\phi_0=(1+k) \, \bigg{\{} V+\left( \frac{f_{4\phi}}{c^4}+\frac{f_{6\phi}}{c^6} \right) \sin 2V \notag \\[1ex]
    &+\left( \frac{g_{4\phi}}{c^4}+\frac{g_{6\phi}}{c^6} \right) \sin 3V+\frac{h_{6\phi}}{c^6} \sin 4V+\frac{i_{6t}}{c^6} \sin 5V\bigg{\}}
\end{align}
where 
\begin{align}
    V=2 \, \tan^{-1} \left( \sqrt{\frac{e_\phi+1}{e_\phi-1}} \, \tanh \frac{u}{2} \right)\,.
\end{align}
\end{subequations}
In the above expressions, 
$k$ and $e_\phi$ denote PN-accurate rate of periastron advance, and certain angular eccentricity, respectively. Additionally,
we have several orbital functions like $f_{4 \phi},~f_{6 \phi},~g_{4\phi},~g_{6 \phi},~i_{6 \phi}$ and $h_{6 \phi}$ that appear at 2PN and 3PN orders.
We note that Ref.~\cite{cho2018gravitational} provided 3PN-accurate expressions 
for various orbital elements and functions in terms of $E$ and $j$.

It is straightforward but tedious to compute 3PN-accurate expressions for $\dot{r}^2, \ v^2$ in terms of $E, \ j^2, \ (e_t \cosh u -1)$.
These dynamical variables that appear in the far-zone energy flux is 
computed by employing the following relations 
	\begin{subequations}
	\begin{align}
	\frac{dt}{du}&=\frac{\partial t}{\partial u}+\frac{\partial t}{\partial v} \, \frac{d v}{d u} \,,\\[0.0ex]
	&\notag\\[0.0ex]
	\dot{r}^2&=\left(\frac{dr}{du} \bigg{/} \frac{dt}{du}\right)^2\,, \\[1ex]
	\dot{\phi}^2&=\left(\frac{d\phi}{dv} \, \frac{dv}{du} \bigg{/}\frac{dt}{du}\right)^2 \,,\\[1ex]
	v^2&=\dot{r}^2+r^2 \, \dot{\phi}^2.
	\end{align}
\end{subequations}
 In what follows, we display 1PN-accurate parametric expressions for $\dot r^2$, $v^2$ 
 and $G\,M/r$ for introducing the reader to the structure of these expressions
\begin{widetext}
 \begin{subequations}
 \label{dotrphi}
\begin{align}
\dot{r^2}_\text{MH}&=\bigg{\{}1+ \frac{2}{(e_t \cosh u -1)}- \frac{(2E j^2)}{(e_t \cosh u -1)^2}\bigg{\}} (2 E) +\frac{(2 E) ^2}{c^2} \bigg{\{}-\frac{3}{4}+\frac{9 \nu }{4}+\frac{1}{(e_t \cosh u -1)^2} \bigg{[}-2+(2E j^2)\nonumber \\[1ex]
& \bigg{(}-\frac{7}{2}-\frac{\nu }{2}\bigg{)}+2 \nu
\bigg{]}+\frac{1}{(e_t \cosh u -1)} \bigg{[}-\frac{3}{2}+\frac{9 \nu }{2}\bigg{]}\bigg{\}} \,,\\[2ex]
v^2_\text{MH}&=\left\{1+ \frac{2}{(e_t \cosh u-1)}\right\} (2 E) +\frac{(2 E)^2}{c^2} \bigg{\{}-\frac{3}{4}+\frac{9 \nu }{4}+ \frac{(2 E j^2) \nu }{(e_t \cosh u-1)^3} +\frac{2 (\nu-1)}{(e_t \cosh u-1)^2}\,, \nonumber \\[1ex]&
+\frac{1}{(e_t \cosh u-1)}\bigg{[}-\frac{3}{2}+\frac{9 \nu }{2}\bigg{]}\bigg{\}}\,,
\end{align}

\begin{align}
\frac{G\,M}{r}&=\frac{2 \, E}{(e_t \, \cosh u-1)}+\frac{(2E)^2}{c^2}\, \bigg{\{} \frac{1}{(1-e_t \, \cosh u)} \left[ \frac{9}{4}-\frac{5 \, \nu}{4}\right]+\frac{1}{(1-e_t \, \cosh u)^2} \, \bigg{[} 4-\frac{3}{2} \, \nu\bigg{]}\bigg{\}}.
\end{align}
\end{subequations}
{
The explicit 3PN-accurate expressions for these dynamical variables are provided at 
\url{https://github.com/subhajittifr/hyperbolic_flux}.
}

\end{widetext}

We are now in a position to replace dynamical variables $r, \dot r$ and $\dot \phi$
that appear in the 3PN-accurate instantaneous far-zone energy flux expression,
given by Eqs.~(5.2) of Ref.~\cite{arun2008inspiralling}, with 
the 3PN-accurate version of the above equations.
The associated 3PN-accurate expression for the 
radiated energy during hyperbolic encounters reads
\begin{align}
\label{DlE_3}
\Delta \mathcal{E}=\int^{+\infty}_{-\infty} dt \, \F=\int^{+\infty}_{-\infty} du \, \left(\frac{dt}{du}\right) \F.
\end{align}
The use of 3PN-accurate expressions for $r, \dot r$ and $\dot \phi$ in terms of 
$E, h,$ and $(1-e_t \, \cosh u)$, as evident from our Eqs.~\eqref{dotrphi}, in the  
Eqs.~(5.2) of Ref.~\cite{arun2008inspiralling} leads to 
\begin{align}
\label{dEdt}
\F&=\frac{1}{n}\,\frac{du}{dt}\,\bigg{\{}\sum_{N=3}^{11}  \frac{\bar{\alpha}_N(E,j)}{(e_t \cosh u-1)^{N}}\notag\\[1ex]
&+\sum_{N=5}^9 \bar{\beta}_N (E,j) 
\frac{\sinh u}{(e_t \cosh u-1)^{N}}\notag \\[1ex]
&+\sum_{N=5}^9 \bar{\gamma}_N (E,j)\,\frac{\ln (e_t \cosh u-1)}{(e_t \cosh u-1)^{N}} \bigg{\}},
\end{align}
where we may write in a compact manner the above constant coefficients as 
\begin{align*}
\bar{W}_N(E,j)=\frac{\nu^2}{G} \left(\frac{E}{c}\right)^5 \, W_N(E,j)\,, 
\end{align*}
where $W_N$ stands for $ \bar \alpha, \bar \beta, \bar \gamma$.
We do not  list all these lengthy coefficients in the manuscript and they 
are provided in  
an ancillary
\textsc{Mathematica} file at (\url{https://github.com/subhajittifr/hyperbolic_flux}).
However, we list below one of them to show the typical structure of these coefficients 
\begin{align}
    \alpha_3(E,j)&=\frac{256}{15}+\frac{E}{c^2} \bigg{(}\frac{13184}{105}-\frac{128 \, \nu }{21}\bigg{)}+\frac{E^2}{c^4}\bigg{(}\frac{51328}{63}\nonumber \\[1ex]
    &+\frac{5504 \, \nu
}{315}+\frac{640 \, \nu ^2}{63}\bigg{)}+\frac{E^3}{c^6} \bigg{(}\frac{5451968}{1155}+\frac{1803328 \nu }{3465}\nonumber \\[1ex]
&+\frac{181792 \, \nu ^2}{3465}-\frac{8000
\, \nu ^3}{693}\bigg{)}
\end{align}
We note that 
 all $\bar{\beta}_N$ terms appear at the 2.5PN order while $\bar{\gamma}_N$ terms are accompanied by the logarithmic terms at the 3PN order.
 Additionally, we have incorporated the dependence  on the constant $\log r_0$
 into the coefficients $\bar{\alpha}_N$. 
 Recall that $r_0$ is the gauge dependent length scale appearing in the definition of source multiple moments~\cite{blanchet2005hadamard} as discussed in Ref.~\cite{arun2008inspiralling}.
 We have indeed  verified that the 1PN-accurate version of these coefficients 
 match with  Ref.~\cite{blanchet1989higher}.
 
 We now move to tackle how these coefficients contribute to the 
 instantaneous 3PN-accurate expression for the radiated energy during hyperbolic 
 encounters.
 It is fairly straightforward to infer that all the 
  2.5PN terms, namely $\bar{\beta}_N$ terms, 
  do not contribute to $\Delta {\cal E}$ expression and this is because
\begin{align}
\int_{-\infty}^\infty  \frac{\sinh u}{(e_t \cosh u-1)^N} \, du=0\,.
\end{align}
This is obviously due to the fact that these integrands are an odd function in $u$.
 The integration of $\bar \alpha_N$ terms should be straightforward as they are essentially similar to the terms that we confronted at the Newtonian order.
Therefore, we employ Eq.~\eqref{int} to compute $\Delta {\cal E}$ contributions that arise from  nine $\bar \alpha_N$ terms in Eq.~\eqref{dEdt}.

Clearly, it is very tricky to  tackle the logarithmic terms 
and we pursued an entirely new line of investigations compared to 
what was done in Ref. \cite{arun2008inspiralling} for the eccentric orbits.
After some detailed efforts, we were able to obtain 
analytic expressions for these integrals though  the final expressions were too lengthy to list here.
However, we were eventually able to obtain the most simplified form of these integrals with the help of Clausen identity as detailed  in 
the Appendix \ref{log_integral}.
This allowed us to tackle the $\bar \gamma_N$ contributions to the 3PN-accurate $\Delta {\cal E}$ expression as 
\begin{align}
\label{log_int}
&\int_{-\infty}^\infty \frac{\ln (e \cosh u -x)}{e \cosh u -x} \, du=\frac{2}{\sqrt{e^2-x^2}} \nonumber \\[1ex]
& \times \bigg{[}\text{Cl}_2 \bigg{(}2 \arccos\bigg{(}-\frac{x}{e}\bigg{)}\bigg{)} +\arccos \bigg{(}-\frac{x}{e}\bigg{)} \nonumber \\[1ex]
& \times \ln \bigg{(}\frac{2(e^2-x^2)}{e}\bigg{)}\bigg{]},
\end{align}
where $\text{Cl}_2(x)$ is the Clausen function of order $2$ given by the integral
\begin{align}
\text{Cl}_2(x)=-\int_0^x dy \,  \ln \left|2 \sin \frac{y}{2}\right| \,.
\end{align}
Interestingly, the Clausen function admits the following Fourier series representation:
\begin{align}
\text{Cl}_2(x)=\sum_{k=1}^\infty \frac{\sin k x}{k^2}\,.
\end{align} 

The crucial integrals that are associated with the ${\bar \gamma_N}$ coefficients 
in Eq.~\eqref{dEdt} can now tackled by noting that 
the integrals $$\int_{-\infty}^\infty \frac{\ln (e \cosh u -1)}{(e \cosh u -1)^N} \, du\,,$$ can be computed after taking successive derivatives of Eq. \eqref{log_int} with respective to $x$ and then finally taking $x \to 1$.

\begin{widetext}

With these inputs and the following 
3PN accurate expression of $E$ in terms of $e_t$ and $j$ \cite{cho2018gravitational},
\begin{align}\label{Eineth}
	&E=\frac{e_t^2-1}{2 \, j^2}+\frac{1}{c^2\,j^4}\bigg{[}-\frac{9}{8}+e_t^2 \bigg{(}\frac{13}{4}-\frac{3 \, \nu }{4}\bigg{)}+e_t^4
		\bigg{(}-\frac{17}{8}+\frac{7 \, \nu }{8}\bigg{)}-\frac{\, \nu }{8}\bigg{]}+\frac{1}{c^4 \, j^6} \bigg{[}-\frac{81}{16}+\frac{7 \, \nu }{16}-\frac{\, \nu ^2}{16}+e_t^4
		\bigg{(}-\frac{339}{16}\nonumber \\[1ex]
		&+\frac{213 \, \nu }{16}-\frac{35 \, \nu ^2}{16}\bigg{)}+e_t^2 \bigg{(}\frac{243}{16}-\frac{29 \, \nu }{16}+\frac{3 \, \nu ^2}{16}\bigg{)}+e_t^6
		\bigg{(}\frac{177}{16}-\frac{191 \, \nu }{16}+\frac{33 \, \nu ^2}{16}\bigg{)}\bigg{]}+\frac{1}{c^6 \, j^8} \bigg{[}-\frac{3861}{128}+\bigg{(}\frac{8833}{384}\nonumber \\[1ex]
		&-\frac{41
		\pi ^2}{64}\bigg{)} \, \nu +\frac{5 \, \nu ^2}{64}-\frac{5 \, \nu ^3}{128}+e_t^6 \bigg{(}\frac{4641}{32}-\frac{19037 \, \nu }{96}+\frac{1119 \, \nu ^2}{16}-\frac{231
		\, \nu ^3}{32}\bigg{)}+e_t^2 \bigg{(}\frac{2105}{32}+\bigg{(}\frac{157553}{3360}-\frac{41 \pi ^2}{32}\bigg{)} \, \nu \nonumber \\[1ex]
	&-\frac{105 \, \nu ^2}{16}+\frac{\, \nu
		^3}{32}\bigg{)}+e_t^4 \bigg{(}-\frac{7479}{64}+\bigg{(}\frac{16123}{2240}+\frac{123 \pi ^2}{64}\bigg{)} \, \nu -\frac{313 \, \nu ^2}{32}+\frac{105 \, \nu
		^3}{64}\bigg{)}+e_t^8 \bigg{(}-\frac{8165}{128}+\frac{15515 \, \nu }{128}\nonumber \\[1ex]
	&-\frac{3435 \, \nu ^2}{64}+\frac{715 \, \nu ^3}{128}\bigg{)}\bigg{]}\,,
\end{align}
we finally obtain the 3PN-accurate instantaneous contributions to $\Delta {\cal E}$ in terms of $e_t$ and $j$,
\begin{align}
\Delta \mathcal{E}&=\frac{2}{15} \,\frac{M \, \nu ^2}{j^7 \, c^5} \Big( \mathcal{I}_N^\text{MH}+\frac{1}{c^2\, j^2}\,  \mathcal{I}_\text{1PN}^\text{MH}+\frac{1}{c^4\, j^4} \,  \mathcal{I}_\text{2PN}^\text{MH}+\frac{1}{c^6\, j^6} \,  \mathcal{I}_\text{3PN}^\text{MH}\Big)\,.
\end{align}
where contributions to $\Delta {\cal E}$ that appear 
	at the Newtonian, 1PN, 2PN and 3PN orders read 
\begin{subequations}
\label{Einst}
\begin{align}
\mathcal{I}_N^\text{MH}&= \sqrt{e_t^2-1}\,\bigg{[}\,\frac{602}{3}+\frac{673 \,e_t^2}{3}\bigg{]} +\arccos\left(-\frac{1}{e_t}\right)\,\bigg{[}96+292\, e_t^2+37 \,e_t^4\bigg{]} 
\,,\\[2ex]
\mathcal{I}_\text{1PN}^\text{MH}&=\sqrt{e_t^2-1}\, \Bigg{[}\frac{153263}{70}-\frac{1547 \, \nu 
}{3}+e_t^2\,
\bigg{(}\frac{271849}{70}-\frac{13799 \, \nu }{6}\bigg{)}+e_t^4\, 
\bigg{(}-\frac{288513}{280}-2 \, \nu \Bigg{)}\Bigg{]}\\[1ex]
&+\arccos\left(-\frac{1}{e_t}\right)
\,\Bigg{[}\frac{6578}{7}-168 \, \nu+e_t^2\,
\bigg{(}\frac{31013}{7}-1982 \, \nu \bigg{)}+e_t^4 \,
\bigg{(}-\frac{223}{4}-\frac{1483 \, \nu }{2}\bigg{)} +e_t^6 \,
\bigg{(}-\frac{15219}{56}+74
\, \nu \bigg{)}\Bigg{]}\,,\notag \\[2ex]
\mathcal{I}_\text{2PN}^\text{MH}&= \sqrt{e_t^2-1} \,
\Bigg{[}\frac{405300022}{19845}-\frac{2947852
	\, \nu }{315}+\frac{1173 \, \nu ^2}{4}+e_t^2 \,
\bigg{(}\frac{6673495637}{158760}-\frac{114248429 \, \nu 
}{2520}+\frac{66217 \, \nu ^2}{8}\bigg{)}\nonumber \\[1ex] 
&+e_t^4\, \bigg{(}-\frac{3823800817}{158760}+\frac{41499527
	\, \nu }{5040}+\frac{3619 \, \nu ^2}{2}\bigg{)}+e_t^6 \,
\bigg{(}\frac{39802111}{7840}-\frac{67328 \, \nu }{35}-103 \, \nu 
^2\bigg{)}\Bigg{]}\notag\\[1ex]
&+\arccos\left(-\frac{1}{e_t}\right) 
\Bigg{[}\frac{1636769}{189}-\frac{74435 \, \nu }{21}+48 \, \nu ^2+e_t^2 \bigg{(}\frac{8638156}{189}-\frac{762901
	\, \nu }{21}+\frac{9851 \, \nu ^2}{2}\bigg{)}\nonumber \\[1ex] 
&+e_t^4 \,
\bigg{(}-\frac{554104}{63}-\frac{350943 \, \nu }{28}+\frac{48063 \, \nu 
	^2}{8}\bigg{)}+e_t^6 \,
\bigg{(}-\frac{1324649}{336}+\frac{611613
	\, \nu }{112}-\frac{1779 \, \nu ^2}{2}\bigg{)}\notag\\[1ex]
	&+e_t^8 \,
\bigg{(}\frac{1224929}{672}-\frac{10070 \, \nu }{7}+185 \, \nu
^2\bigg{)}\Bigg{]}\,,\\[2ex]
\mathcal{I}_\text{3PN}^\text{MH}&= \sqrt{e_t^2-1} \, \Bigg{[}\frac{6713608}{1575}+\frac{17868572 \,  
e_t^2}{525}  +\frac{19300553 \, e_t^4}{525}+\frac{17525209  \,
	e_t^6}{3150}\Bigg{]} \, \log \left[ \frac{2\,r_0\,(e_t^2-1)}{G\,M \, j^2 \, e_t} \right] -\Bigg{[}\frac{54784}{35}+\frac{465664 \, e_t^2}{21}\nonumber \\[1ex] 
	&+\frac{4426376 \, 
	e_t^4}{105}+\frac{1498856 \, e_t^6}{105}+\frac{31779 \, 
	e_t^8}{70}\Bigg{]} \, 
\Bigg{\{} \log \left[ \frac{2\,G\,M\,  \, j^2}{e_t\,r_0 } \right] \, \arccos \left(-\frac{1}{e_t} \right) +\text{Cl}_2 \left[ 2\, \arccos \left(-\frac{1}{e_t} \right) \right]\Bigg{\}}\nonumber \\[1ex]
&+\sqrt{e_t^2-1} \, 
\Bigg{[}\frac{1959816183329}{8731800}+\Bigg{(}-\frac{69960810223}{317520}+\frac{11632643\, \pi ^2}{3360}\Bigg{)} \, \nu +\frac{9109459 \, \nu ^2}{560}+\frac{689
	\, \nu ^3}{24}\nonumber \\[1ex] 
&+e_t^2 
\Bigg{(}\frac{11238026145523}{17463600}+\Bigg{(}-\frac{264708911281}{317520}+\frac{998227 \, 
	\pi ^2}{140}\Bigg{)}
\, \nu +\frac{518878433 \, \nu ^2}{2520}-\frac{641483 \, \nu 
	^3}{48}\Bigg{)}
\nonumber \\[1ex] 
&+e_t^4 \, \Bigg{(}-\frac{10459843311391}{139708800}+\Bigg{(}\frac{28976695225}{254016}+\frac{99671 \,
	\pi ^2}{896}\Bigg{)} \, \nu+\frac{143029027
	\, \nu ^2}{2880}-16908 \, \nu ^3\Bigg{)}+e_t^6 \,
\Bigg{(}\frac{135148514527}{739200}\nonumber \\[1ex]
&+\Bigg{(}-\frac{15436846447}{211680}-\frac{12303239 \, \pi ^2}{13440}\Bigg{)}
\, \nu +\frac{12729151 \, \nu ^2}{5040}+\frac{8501 \, \nu^3}{3}\Bigg{)}+e_t^8 
\Bigg{(}-\frac{112472361473}{4139520}\nonumber \\[1ex]
&+\frac{244676087 \, \nu 
}{8820}-\frac{432849
	\, \nu ^2}{140}-\frac{1147 \, \nu ^3}{3}\Bigg{)} \Bigg{]} +\arccos\left(-\frac{1}{e_t}\right) \Bigg{[}\,  \frac{20510192533}{207900}+\Bigg{(}-\frac{69631105}{756}+1599 \pi
^2\Bigg{)} \, \nu \nonumber \\[1ex] 
&+\frac{145195 \, \nu ^2}{28}+9 \, \nu ^3+e_t^2 \,\Bigg{(}\frac{269134761733}{415800}+\Bigg{(}-\frac{736708039}{1080}+\frac{55637 \, \pi ^2}{8}\Bigg{)} \, \nu +\frac{21177007 \, \nu ^2}{168}-\frac{22283 \, \nu 
  ^3}{4}\Bigg{)}\notag \\[1ex]
&+e_t^4 \,\Bigg{(}\frac{125593677691}{554400} +\Bigg{(}-\frac{2952969469}{10080}+\frac{172405\, \pi^2}{64}\Bigg{)} \, \nu +\frac{112045205 \, \nu ^2}{672}-\frac{381255 \, \nu^3}{16}\Bigg{)}+e_t^6 \, 
\Bigg{(}\frac{115347955537}{1108800}\notag \\[1ex]
&+\Bigg{(}\frac{7251474763}{60480} 
-\frac{174619 \, \pi^2}{128}\Bigg{)} \, \nu-\frac{2843351 \, \nu ^2}{64}+3155 \, \nu^3\Bigg{)}+e_t^8 \, \Bigg{(}\frac{22565112667}{492800}+\Bigg{(}-\frac{37470739}{672}-\frac{12177 \, \pi ^2}{128}\Bigg{)} \, \nu \notag \\[1ex]
&+\frac{2686573 \, \nu ^2}{112} -2075 \, \nu^3\Bigg{)}+e_t^{10} \,  \Bigg{(}-\frac{484326439}{39424}+\frac{502175 
  \, \nu }{28}-\frac{25061
  \, \nu ^2}{4}+518 \, \nu ^3\Bigg{)} \Bigg{]}\,.
	\end{align}
\end{subequations}
We have found the full agreement with Eq.(1.4) of Ref.~\cite{Herrmann:2021lqe} upto 3PN($v^7$) order in bremsstrahlung limit. Note that since the choice of the relation between $e_t$, $E$ and $j$ (Eq.~\ref{Eineth}) was made in favor of convenience of computation under MH gauge condition \cite{cho2018gravitational}, hence the apparent expression of Eq.~(\ref{Einst}) (and also Eq.~(\ref{Jinst})) is not gauge invariant. 
We have verified that the above expression is fully consistent with Eqs.~(C9) of Ref.~\cite{bini2021radiative} up-to 2PN order which required us to express $e_t$ and $j$ in terms of $e_r$ and $a_r$ to 2PN order in the MH gauge.
\par We note that it is customary to characterize hyperbolic encounters with the help of an impact parameter $b$ and an eccentricity parameter as noted in Ref.~\cite{junker1992binary}.
Therefore, we provide 3PN accurate relation that connects $j$ to $b$ and 
$e_t$ with the help of Ref.~\cite{cho2018gravitational}.
\begin{align}
j^2&=\frac{b}{G \, M} \, \sqrt{e_t^2-1} \,  \bigg{\{}1+\bigg{[}1+\frac{\, \nu }{6}+e_t^2 \bigg{(}-4+\frac{17 \, \nu }{6}\bigg{)}\bigg{]} \rho +\bigg{[}-\frac{17}{4}-\frac{61
		\, \nu }{24}-\frac{\, \nu ^2}{8}+e_t^4 \bigg{(}-\frac{49}{4}+\frac{307 \, \nu }{24}-\frac{95 \, \nu ^2}{24}\bigg{)}\nonumber \\[1ex] 
&+e_t^2 \bigg{(}8+\frac{23 \, \nu
	}{4} -\frac{5 \, \nu ^2}{12}\bigg{)}\bigg{]} \rho^2 +\bigg{[}-\frac{23}{4}+\bigg{(}\frac{1375}{48}-\frac{41 \pi ^2}{64}\bigg{)} \, \nu -\frac{2213
	\, \nu ^2}{72}+\frac{\, \nu ^3}{12}+e_t^2 \bigg{(}-\frac{125}{2}+\bigg{(}\frac{141199}{1680}\nonumber \\[1ex] 
&-\frac{123 \pi ^2}{64}\bigg{)} \, \nu +\frac{607 \, \nu ^2}{24}+\frac{5
	\, \nu ^3}{4}\bigg{)}+e_t^4 \bigg{(}\frac{333}{4}-\frac{557 \, \nu }{16}-\frac{595 \, \nu ^2}{24}+\frac{95 \, \nu ^3}{12}\bigg{)}+e_t^6 \bigg{(}-\frac{271}{3}+\frac{7339
	\, \nu }{48}-\frac{6391 \, \nu ^2}{72}\nonumber \\[1ex]
&+\frac{217 \, \nu ^3}{12}\bigg{)}\bigg{]} \rho ^3 \bigg{\}}\,, 
\end{align}
where $\rho=\frac{1}{\sqrt{e_t^2-1}}\, \frac{G \, M}{ b \, c^2}\,.$
 We now move on the briefly list our approach to compute PN-accurate the radiated angular
 momentum during hyperbolic encounters that extends the 1PN-accurate effort of Ref.~\cite{junker1992binary,bini2021radiative}.
\end{widetext}

\subsection{3PN-accurate instantaneous contributions to the radiated angular momentum}

\label{am_3pn}
   The crucial input that is required for our $\Delta {\cal J}$ 
   computation is the 3PN-accurate instantaneous contributions to
   the far-zone GW angular momentum flux from compact binaries in non-circular orbits, given by 
   Eqs.~(3.4) in Ref.~\cite{arun2009third} and therefore in the 
   MH gauge.
   The dynamical variables that appear in these PN-contributions that includes $|\mathbf{\tilde{L}_N}|$  
   are expressed in terms of $E,j$ and $(e_t\, \cosh u -1)$ to 3PN order with the help of Ref.~\cite{cho2018gravitational}. 
The resulting 3PN extension of Eq. \eqref{G_N} 
that provides 3PN-accurate instantaneous contributions to
   the scalar far-zone GW angular momentum flux
may be written as 
	\begin{align}
	\Delta\mathcal{J}& =\frac{1}{n}\,\frac{du}{dt}\,\bigg{\{}\sum_{N=2}^{10}  \frac{\bar{A}_N(E,j)}{(e_t \cosh u-1)^{N}}\nonumber \\[1ex]
	&+\sum_{N=4}^8 \bar{B}_N (E,j) 
	\,\frac{\sinh u}{(e_t \cosh u-1)^{N}}\notag\\[1ex]
	&+\sum_{N=4}^8 \bar{\Gamma}_N (E,j) 
	\frac{\ln (e_t \cosh u-1)}{(e_t \cosh u-1)^{N}} \bigg{\}}\,,
	\end{align}
	where
	\begin{align*}
	\bar{Y}_N (E,j)=\frac{\nu^2\,E^4\,M\,j}{c^5}\,Y_N (E,j)\,,
	\end{align*}
where $Y$ stands for $A,B,\Gamma$.
It is obvious that we can pursue the similar arguments,
detailed in Sec.~\ref{inst_en},
for computing 
 $\int^{+\infty}_{-\infty}\, dt\,\G(j, e_t, u) $.
This leads to the following 3PN-accurate instantaneous contributions to the radiated angular momentum during hyperbolic encounters of 
non-spinning compact objects 
\begin{widetext}
\begin{align}
\Delta \mathcal{J}&=\frac{8}{5 \, c^5} \, \frac{G M^2\, \nu^2}{ j^4}\,  \Big( \mathcal{H}_N^\text{MH}+\frac{1}{c^2\, j^2}\,  \mathcal{H}_\text{1PN}^\text{MH}+\frac{1}{c^4\, j^4} \,  \mathcal{H}_\text{2PN}^\text{MH}+\frac{1}{c^6\, j^6}\,  \mathcal{H}_\text{3PN}^\text{MH}\Big),
\end{align}
where the individual contributions that appear at Newtonian, 1PN, 2PN and 3PN orders are given by 
\begin{subequations}
\label{Jinst}
\begin{align}
    \mathcal{H}^\text{MH}_\text{N}&=\sqrt{e_t^2-1} \, \Bigg{(}13+2 
e_t^2\Bigg{)} +\arccos \left(-\frac{1}{e_t}\right) \, \Bigg{(}8+7 e_t^2\Bigg{)} 
 \,, \\[2ex]
  \mathcal{H}^\text{MH}_\text{1PN}&=
  \sqrt{e_t^2-1} 
\Bigg{[}\frac{14759}{168}+e_t^2 \,
\Bigg{(}\frac{11153}{336}-\frac{1975
	\, \nu }{36}\Bigg{)}-\frac{847 \, \nu }{18}+e_t^4 \,  \Bigg{(}-\frac{62}{7}+4 
\, \nu \Bigg{)}\Bigg{]}+\arccos \left(-\frac{1}{e_t}\right) \Bigg{[}\frac{1777}{42}  \\[1ex]
&+e_t^2 \, \Bigg{(}\frac{3649}{42}-\frac{241
	\, \nu }{3}\Bigg{)}+e_t^4 \,  \Bigg{(}-\frac{5713}{336}+\frac{5 \, \nu}{12}\Bigg{)}-18 \, \nu \Bigg{]} \,,\notag \\[2ex]
\mathcal{H}^\text{MH}_\text{2PN} &=\sqrt{e_t^2-1} \Bigg{[}\frac{38409857}{68040}-\frac{3668237
	\, \nu }{7560}+\frac{785 \, \nu ^2}{18}+e_t^2 \,
\Bigg{(}\frac{2630029}{17010}-\frac{13151567 \, \nu }{15120}
+\frac{4423 \, \nu^2}{12}\Bigg{)}+e_t^4 \,  \Bigg{(}-\frac{853417}{3360}\notag \\[1ex]
&+\frac{1691057 \, \nu }{5040}-\frac{617 \, \nu 
	^2}{18}\Bigg{)}+e_t^6 \, \Bigg{(}\frac{2480}{63}-\frac{755
	\, \nu }{14}+10 \, \nu ^2\Bigg{)}+\Bigg{]}+\arccos 
\left(-\frac{1}{e_t}\right) \,
\Bigg{[}\frac{326917}{1134}-\frac{27031
	\, \nu }{126}+10 \, \nu ^2 \notag \\[1ex]
&+e_t^2 \, \Bigg{(}\frac{414821}{756}-\frac{29055 \, \nu }{28}+298 \, \nu 
^2\Bigg{)}+e_t^4 \,
\Bigg{(}-\frac{1154087}{3024}+\frac{72917
	\, \nu }{336}+\frac{941 \, \nu ^2}{12}\Bigg{)}+e_t^6\, \Bigg{(}\frac{99103}{2016}-\frac{12847 
	\, \nu }{336}+\frac{3 \, \nu ^2}{2}\Bigg{)}\Bigg{]} \,,\notag \\[2ex]
\mathcal{H}^\text{MH}_\text{3PN}&=\sqrt{e_t^2-1} \,  \Bigg{[}\frac{99724}{315}+\frac{351067
	\, e_t^2}{315}+\frac{210683 \, e_t^4}{630}\Bigg{]} \, \log \left[ \frac{2\,r_0\, (e_t^2-1)}{G\,M \, j^2 \, \, e_t} \right] -\Bigg{[}\frac{13696}{105}+\frac{98012 \, e_t^2}{105}+\frac{23326 
	\, e_t^4}{35}+\frac{2461 \, e_t^6}{70}\Bigg{]} \notag \\[1ex]
& \times \, \Bigg{\{} \log \left[ \frac{2 \, G\,M \, j^2}{r_0\, e_t } \right] \, \arccos \left(-\frac{1}{\, e_t} \right) +\text{Cl}_2 \left[ 2\, \arccos \left(-\frac{1}{\, e_t} \right) \right]\Bigg{\}}
+\sqrt{\, e_t^2-1} \, 
\Bigg{[}\frac{55475721271}{8382528}+\Bigg{(}-\frac{17854035221}{1905120}\notag \\[1ex]
&+\frac{313363 \,
	\pi ^2}{1920}\Bigg{)} \, \nu +\frac{636197 \, \nu ^2}{540}-\frac{103 \, \nu 
	^3}{18}+e_t^2 \,  
\Bigg{(}\frac{550589812147}{83825280} 
+\Bigg{(}-\frac{49917375859}{3810240}+\frac{541733
	\pi ^2}{3840}\Bigg{)} \, \nu \notag \\[1ex]
&+\frac{12681271 \, \nu ^2}{1890} -\frac{33883 \, \nu 
	^3}{36}\Bigg{)}+e_t^4 \,  
\Bigg{(}-\frac{276385167053}{335301120}+\Bigg{(}\frac{11543781001}{1905120}
-\frac{22427
	\pi ^2}{1920}\Bigg{)} \, \nu -\frac{116779321 \, \nu ^2}{60480}\notag \\[1ex]
&-\frac{21775 
	\, \nu ^3}{144}\Bigg{)} +e_t^6 \,  
\Bigg{(}\frac{12794620753}{8279040}-\frac{197812189
	\, \nu }{70560}+\frac{9014755 \, \nu ^2}{8064}-\frac{23497 \, \nu 
	^3}{288}\Bigg{)} 
+\, e_t^8 \Bigg{(}-\frac{13784}{77}+\frac{2999 \, \nu 
}{6}\notag \\[1ex]
&-254 \, \nu ^2+28
\, \nu ^3\Bigg{)}\Bigg{]}+\arccos \left(-\frac{1}{\, e_t}\right) 
\Bigg{[}\frac{4577461991}{1247400}+\Bigg{(}-\frac{21428779}{4536}+\frac{369 \pi^2}{4}\Bigg{)} \, \nu 
+\frac{7853 \, \nu ^2}{18}\notag \\[1ex]
&+e_t^2 \, 
\Bigg{(}\frac{13811878057}{1247400}
+\Bigg{(}-\frac{86352541}{5670} +\frac{5781 
	\pi ^2}{32}\Bigg{)} \, \nu +\frac{646651 \, \nu ^2}{126}-488 \, \nu 
^3\Bigg{)}+\, e_t^4 \, 
\Bigg{(}\frac{602403517}{831600}+\Bigg{(}\frac{150579449}{60480}\notag \\[1ex]
&+\frac{6273 \pi^2}{256}\Bigg{)} \, \nu +\frac{21055 \, \nu ^2}{16}-\frac{4289 \, \nu 
	^3}{6}\Bigg{)} + e_t^6 \, \Bigg{(}\frac{2135052803}{1108800}+\Bigg{(}-\frac{1094353}{672}-\frac{615
	\pi ^2}{128}\Bigg{)} \, \nu +\frac{8513 \, \nu ^2}{168}+\frac{565 \, \nu^3}{12}\Bigg{)}\notag \\[1ex]
&+ e_t^8 \, 
\Bigg{(}-\frac{94124017}{709632}+\frac{757831 \, \nu }{2016}-\frac{306977 \, \nu^2}{2688}+\frac{129 \, \nu ^3}{32}\Bigg{)} \Bigg{]}.
	\end{align}
\end{subequations}
\end{widetext}
Note that the first line contributions in 
$\mathcal{H}^\text{MH}_\text{3PN} $ are due to the log terms in the far-zone angular momentum flux. We have verified that our expressions are consistent with Eqs. (E6)
of  Ref.~\cite{bini2021radiative} at the 2PN order. 
We now explain why our instantaneous results can be 
treated to be exact up to 3PN order.

\subsection{On the exact nature of our 3PN results}\label{general}

We note that two crucial inputs are required to compute 
3PN-accurate  expressions for the instantaneous contributions to $\Delta {\cal E} $
and $\Delta {\cal J}$.
The first input is the 3PN-accurate generalized 
quasi-Keplerian parametric solution for compact binaries 
in hyperbolic orbits \cite{cho2018gravitational}.
This solution, presented in 
the 
modified harmonic(MH) gauge, 
provided analytic expressions
for the angular and radial dynamical variables of
the 3PN accurate conservative dynamics of compact binaries in hyperbolic orbits.
With the help of Ref.~\cite{cho2018gravitational}, 
we write schematically 
analytic expressions for these dynamical variables 
as 
\begin{align}
\begin{split}
\label{input_par}
r&=\mathcal{R}(j, \,e_t,\, u)\,,\\[0.5ex]
\phi&=\mathcal{P}(j, \,e_t,\, u)\,,\\[0.5ex]
\dot{r}&=\mathcal{S}(j, \,e_t,\, u)\,,\\[0.5ex]
\dot{\phi}&=\mathcal{Q}(j, \,e_t,\, u)\,,
\end{split}
\end{align}
where $r$ and $\dot r$ stand for the radial orbital 
separation and its time derivative. Further, 
$\phi$ denotes the angular variable of the reduced mass
$\mu$ around the total mass $M$ while $\dot \phi$ is the time derivative of the above orbital phase.
For the present discussion, we employed certain time eccentricity $e_t$
and the reduced angular momentum 
as orbital parameters 
to characterize the hyperbolic orbit.
The temporal evolution arises via the eccentric anomaly
$u$ and it is related to coordinate time $t$ 
via the PN-accurate Kepler equation that we symbolically write as
\begin{align}
\label{inpus_par2}
t&=\mathcal{T}(j, \,e_t,\, u)\,.
\end{align}
Let us  emphasize that above parametric solution 
incorporates only the conservative temporal evolution of orbital  variables to the 3PN order.

The second crucial ingredient for our present computation 
is the 3PN-accurate instantaneous contributions to the energy and angular momentum fluxes, given by Eqs.~(5.2) of Ref.~\cite{arun2008inspiralling} and Eqs.~(3.4) of Ref.~\cite{arun2009third}, in the MH gauge.
For the present discussion, we write these fluxes 
Schematically as 
\begin{align}
\label{inputs_flux}
\begin{split}
\F&=\F(r\,,\dot{r}\,,\dot{\phi})\,,\\[1ex]
\G&=\G(r\,,\dot{r}\,,\dot{\phi})\,.
\end{split}
\end{align}

Following Refs.~\cite{blanchet1989higher,junker1992binary},
we estimate the radiated energy and angular momentum during hyperbolic encounters by integrating the above fluxes from $t=-\infty$ to $t=\infty$.
The analytic treatment of these integrals require us to express the dynamical variables that appear in Eqs.~\eqref{input_par} and Eqs.~\eqref{inpus_par2} with the help of  PN-accurate 
Keplerian parametric solution of Ref.~\cite{cho2018gravitational}.
This leads to 
\begin{subequations}\label{delEJ}
\begin{align}
\Delta \mathcal{E}:=\int^{+\infty}_{-\infty}\, dt\,\F(j, e_t, u) \,,\\[1ex]
\Delta \mathcal{J}:=\int^{+\infty}_{-\infty}\, dt\,\G(j, e_t, u) \,.
\end{align}
\end{subequations}
The above approach is appropriate as it is customary to write these fluxes as 
$\F=-\frac{d\mathcal{E}}{dt}$ and $\G=-\frac{d\mathcal{J}}{dt}$.
However, the far-zone energy and angular momentum fluxes and the time derivatives of 
orbital energy and angular momentum  are related to each other modulo certain total time derivatives that appear at the 2.5PN order \cite{iyer1993post,iyer1995post,gopakumar1997second}.
 This is why the above equalities hold in an orbital averaged sense in the case of bound elliptical orbits \cite{blanchet1989higher}.  

 When we pursue the computations of $\Delta {\cal E}$ and $\Delta {\cal J}$ to 3PN order, there are certain subtleties that we need to address.
 This is related to the fact that both $h$ and $e_t$ vary with time due to the gravitational radiation reaction effects that appear at the 2.5PN order.
 This implies that the temporal evolution in the above integrands occur not only through $u$ but also through $j$ and $e_t$.
 However, the perturbative nature of GW emission allows 
 us to write 
\begin{align}
\begin{split}\label{toypert}
e_t(t) &=e_{t0}+\frac{\delta e_t(t)}{c^5}+\mathcal{O}(c^{-7})\,,\\[1ex]
j(t) &=j_{0}+\frac{\delta j(t)}{c^5}+\mathcal{O}(c^{-7})\,,
\end{split}
\end{align}
where $e_{t0}$ and $j_{0}$ are the values of time eccentricity and the scaled angular momentum at periastron
(defined by $u=0$)  and hence constants.
Therefore, we could ignore temporal evolution in $e_t$ and $j$ 
while computing 
$\Delta {\cal E}$ and $\Delta {\cal J}$ expressions upto 2PN order. 
Further, the resulting expressions involve only the constant scaled orbital 
angular momentum and time-eccentricity along with the two mass parameters $m_1$
and $m_2$.
However, a close look of Eqs.~\eqref{toypert} and its implications for the integrands of 
Eqs.~\eqref{delEJ} 
reveal that 
the dissipative corrections $\delta e_t$, $\delta j$ are required if  we plan to 
obtain 3PN extension of Refs.~\cite{blanchet1989higher,junker1992binary}.

This mainly arises due to the structure of the relative acceleration at 2.5PN order which may be written as $\ddot {\bf x}= {\bf a}_{2.5PN}( r, \dot r, \dot \phi, {\bf r}, {\bf v}) $ and 
this ensures both $ d {\cal E}/dt \neq 0; d {\cal J}/dt \neq 0$ at the 2.5PN order 
and hence contain terms of ${\cal O}(1/c^5)$.
Therefore, we may try to parametrise the orbital dynamics at 2.5PN in the following 
manner
\begin{align}
\begin{split}
r&=\mathcal{R}_0+\frac{1}{c^5}\partial\mathcal{R}_0\cdot \delta\,,\\[0.5ex]
t&=\mathcal{T}_0+\frac{1}{c^5}\partial\mathcal{T}_0\cdot \delta+\frac{1}{c^5}\,C_t(t)\,,\\[0.5ex]
\phi&=\mathcal{P}_0+\frac{1}{c^5}\partial\mathcal{P}_0\cdot \delta+\frac{1}{c^5}\,C_\phi(t)\,,\\[0.5ex]
\dot{r}&=\mathcal{S}_0+\frac{1}{c^5}\partial\mathcal{S}_0\cdot \delta\,,\\[0.5ex]
\dot{\phi}&=\mathcal{Q}_0+\frac{1}{c^5}\partial\mathcal{Q}_0\cdot \delta\,,
\end{split}
\end{align}
where we used a few short hand notation such that 
\begin{align*}
\mathcal{R}_0&=\mathcal{R}(j_0, e_{0t}, u)\,,\\[0.5ex]
\partial\mathcal{R}_0&=\{\frac{\partial }{\partial j}\mathcal{R}(j_0, e_{0t}, u),\frac{\partial }{\partial e_t}\mathcal{R}(j_0, e_{0t}, u)\}\,,\\[0.5ex]
\delta&=\{\delta j, \delta e_t\}\,,
\end{align*}
 and similar notational conventions apply for the other dynamical variables.
 The above expressions are influenced by the 
  improved `method of variation of constants', detailed in Ref.~\cite{damour2004phasing} 
  that provided a way to include the effects of quadrupolar GW emission 
  on the 2PN-accurate Keplerian type parametric solution for eccentric compact
  binaries. Further, the two new variables 
$C_t$ and $C_\phi$  that appear at the 2.5PN order are influenced by 
the $c_l$ and $c_m$ variables of Ref.~\cite{damour2004phasing}.
  It is not difficult to argue that the temporal evolution of these new variables 
should follow 
\begin{align*}
\frac{d C_t}{dt}&=\delta\cdot\Big(\frac{\partial_u\partial\mathcal{R}_0}{\partial_u\mathcal{R}_0}-\frac{\partial_u\partial\mathcal{T}_0}{\partial_u\mathcal{T}_0}-\frac{\partial\mathcal{S}_0\,\partial_u\mathcal{T}_0}{\partial_u\mathcal{R}_0}\Big)\\[0.5ex]
&+\frac{d\delta}{dt}\cdot\Big(-\partial\mathcal{T}_0+\frac{\partial\mathcal{R}_0\,\partial_u\mathcal{T}_0}{\partial_u\mathcal{R}_0}\Big)\,,\\[1ex]
\frac{d C_\phi}{dt}&=\frac{d\delta}{dt}\cdot(-\partial\mathcal{P}_0+\frac{\partial\mathcal{R}_0\,\partial_u\mathcal{P}_0}{\partial_u\mathcal{R}_0})\\[0.5ex]
+\delta\cdot&\Big(\partial\mathcal{Q}_0-\frac{\partial\mathcal{S}_0\,\partial_u\mathcal{P}_0}{\partial_u\mathcal{R}_0}-\frac{\partial_u\partial\mathcal{P}_0}{\partial_u\mathcal{T}_0}+\frac{\partial_u\mathcal{P}_0\partial_u\partial\mathcal{R}_0}{\partial_u\mathcal{R}_0\partial_u\mathcal{T}_0}\Big)\,.
\end{align*}
The fact that ${\bf a}_{\rm 2.5PN}$ does not explicitly depend on $\phi$ and $t$ 
ensure that $C_t$ and $C_\phi$ can not contribute to the variations in 
${\cal E}$ and ${\cal J}$ at 2.5PN order.
In other words, we may write the total time derivative of the conserved energy 
at 2.5PN to be 
\begin{align}
\frac{d\mathcal{E}}{dt}=
\dot{ \mathcal{E}}_0 +  \delta\cdot\partial\dot {\mathcal{E}}_0\,,
\end{align}
where $ \dot {\mathcal{E}}_0$ stands for 2.5PN accurate energy flux, evaluated 
at the periastron $e_t=e_{t0}$ and $j=j_0$.
It is easily seen that $\delta\cdot\partial\dot {\mathcal{E}}_0$ represents radiation reaction correction to leading order radiation, which is induced by the deflection. Note that  $ \dot {\mathcal{E}}_0$ is even function in time, because the choice of $z$ axis (perpendicular to the orbital plane) in the opposite way should be formally equivalent to time reverse operation $t\leftrightarrow -t$, which cannot make any difference in the result and in other words, there is no radiation reaction in $\dot {\mathcal{E}}_0$).  Further, $\partial\dot {\mathcal{E}}_0$ is also even in that they are just partial derivatives of $ \dot {\mathcal{E}}_0$ with respect to the $e_t, \,j$, so they can not affect to its time dependency structure. On the other hand, $\delta(t)$ comes from time integration of $\dot {\mathcal{E}}_0$ and therefore it should be odd in time. 
We may now conclude that the reaction correction $\delta\cdot\partial\dot {\mathcal{E}}_0$ is also odd function. Hence, when it comes to total radiation, which involves the integration from $t=-\infty$ to $t=+\infty$, $\delta\cdot\partial\dot {\mathcal{E}}_0$ does not contribute to $\Delta \mathcal{E}$ (likewise to $\Delta \mathcal{J}$) at all.\\

These arguments ensure that we can express the various PN contributions to the $\Delta \mathcal{E}$ and $\Delta \mathcal{J}$ integrands in terms of variables that are associated 
 with the Keplerian type solution to the PN-accurate conservative dynamics of hyperbolic encounters.
 In other words, we are fully justified to ignore Newtonian order GW emission induced 
 variations in $\Delta \mathcal{E}$ and $\Delta \mathcal{J}$  while characterizing the orbital 
 dynamics during the 3PN order  $\Delta \mathcal{E}$ and $\Delta \mathcal{J}$  computations.
 Therefore, we employ $e_t$ and $j$ to characterize PN-accurate hyperbolic orbits 
 and can treat them as constant parameters during computing 3PN-accurate expressions for the radiated energy and angular momentum.
This is why we have not considered the radiation reaction while we computed the 3PN accurate instantaneous radiation, and hence this argument fully completes our computation. 

\subsection{Parabolic limit}

Here, we list the exact values of the parabolic limit $e_t=e=1$, or $E=0$ of the total radiations.


\begin{align}
&\Delta\mathcal{E}(E=0) =\frac{M \, \nu^2}{15}\Bigg[ \frac{850 \pi }{j^7}+\frac{7 \pi  (5763-3220 \, \nu )}{4 j^9}\notag\\[0.5ex]
&+\frac{\pi  (29198255+378 \, \nu  (-86017+18270 \, \nu ))}{336 \, j^{11}}\\[0.5ex]
&+\frac{\pi}{80640 \, j^{13}} \Big(179020439969+440
   \nu  \big(-361091813\notag\\[0.5ex]
   &+3587787 \pi ^2+945 (105219-10780 \, \nu ) \, \nu \big)\notag\\[0.5ex]
&-13003119360 \log \left(2 \,\frac{G\,M}{r_0}\, j^2\right)\Big)\Bigg]\,,\notag\\[1ex]
&\Delta\mathcal{J}(E=0) =\frac{8\,G M^2 \, \nu^2}{5}\Bigg[ \frac{15 \pi }{j^4}-\frac{5\, \pi  \,(-1077+940 \, \nu )}{48 j^6}\notag\\[0.5ex]
&+\frac{\pi \, \left(1307683-2782332 \,\nu +1005480\, \nu ^2\right)}{2592 \, j^8}\\[0.5ex]
&+\frac{1}{322560 \, 
   j^{10}}\Big(\pi \,
   \big(5567205457+20 (-301863524\notag\\[0.5ex]
&+4719141 \, \pi ^2) \, \nu+2200128840 \, \nu ^2-371498400 \, \nu ^3\big)\notag\\[0.5ex]
&-569479680 \, \pi  \log \left(2 \,\frac{G\,M}{r_0}\, j^2\right)\Big)\Bigg]\,.\notag
\end{align}
We have verified that these expressions are in full agreement 
with the parabolic limit of 
the radiated energy and angular momentum during one radial period of an eccentric binary available in Refs.~\cite{arun2008inspiralling, arun2009third}.

\subsection{Implications of Post-Bremsstrahlung Expansion}

We now probe the implications of the post-bremsstrahlung limit
of our $\Delta {\cal E}$ and $\Delta {\cal J}$ expressions.
Recall that  the bremsstrahlung limit arises by 
allowing the eccentricity parameter $\rightarrow \infty$ as noted in Ref.~\cite{blanchet1989higher}, and we are exploring the implications 
of eccentricity corrections to such 
bremsstrahlung limit of our 3PN order hyperbolic expressions.
This effort is also influenced by the fact that it is rather 
difficult to obtain closed form expressions for 
$\Delta {\cal E}$ and $\Delta {\cal J}$ expressions even when the leading order hereditary contributions are included \cite{bini2021higher}.
Therefore, it is reasonable that our ongoing effort to obtain fully 3PN-accurate expressions for the radiated energy and angular momentum 
in GWs during hyperbolic encounters will not be exact in orbital 
eccentricity.
In what follows, we probe the implications of 
post-bremsstrahlung  expansion of our 3PN order hyperbolic
$\Delta {\cal E}$ and $\Delta {\cal J}$ expressions  
with respect to their bound orbit counterparts.
These counterparts are essentially 3PN-accurate instantaneous 
$\delta {\cal E}$ and $\delta {\cal J}$ expressions that provide 
the radiated energy and angular momentum during one radial period of an eccentric binary.
These expressions can easily be obtained from
Refs.~\cite{arun2008inspiralling, arun2009third}, 
and are exact in orbital eccentricity.
Further, we find it convenient to 
employ the Newtonian eccentricity parameter $e=\sqrt{1+2 \, E \, j^2}$ to characterize the both the bound and unbound far-zone quantities  to ensure that the same eccentricity parameter is used in our comparisons.
We display the relevant $\Delta {\cal E}$ and $\Delta {\cal J}$ expressions in Appendix.~\ref{appendix_AC2}.
Further, 
the explicit 3PN-accurate instantaneous 
post-bremsstrahlung $\Delta \cal E$ and $\Delta \cal J$ expressions for hyperbolic encounters are available at 
\url{https://github.com/subhajittifr/hyperbolic_flux} \, ,where $\Delta \cal E$ is expanded up-to $\mathcal{O}(1/{e^5})$ while $\Delta \cal J$ has been expanded up-to $\mathcal{O}(1/{e^7})$.

\begin{figure}[h]
    \centering
    \includegraphics[scale=0.6]{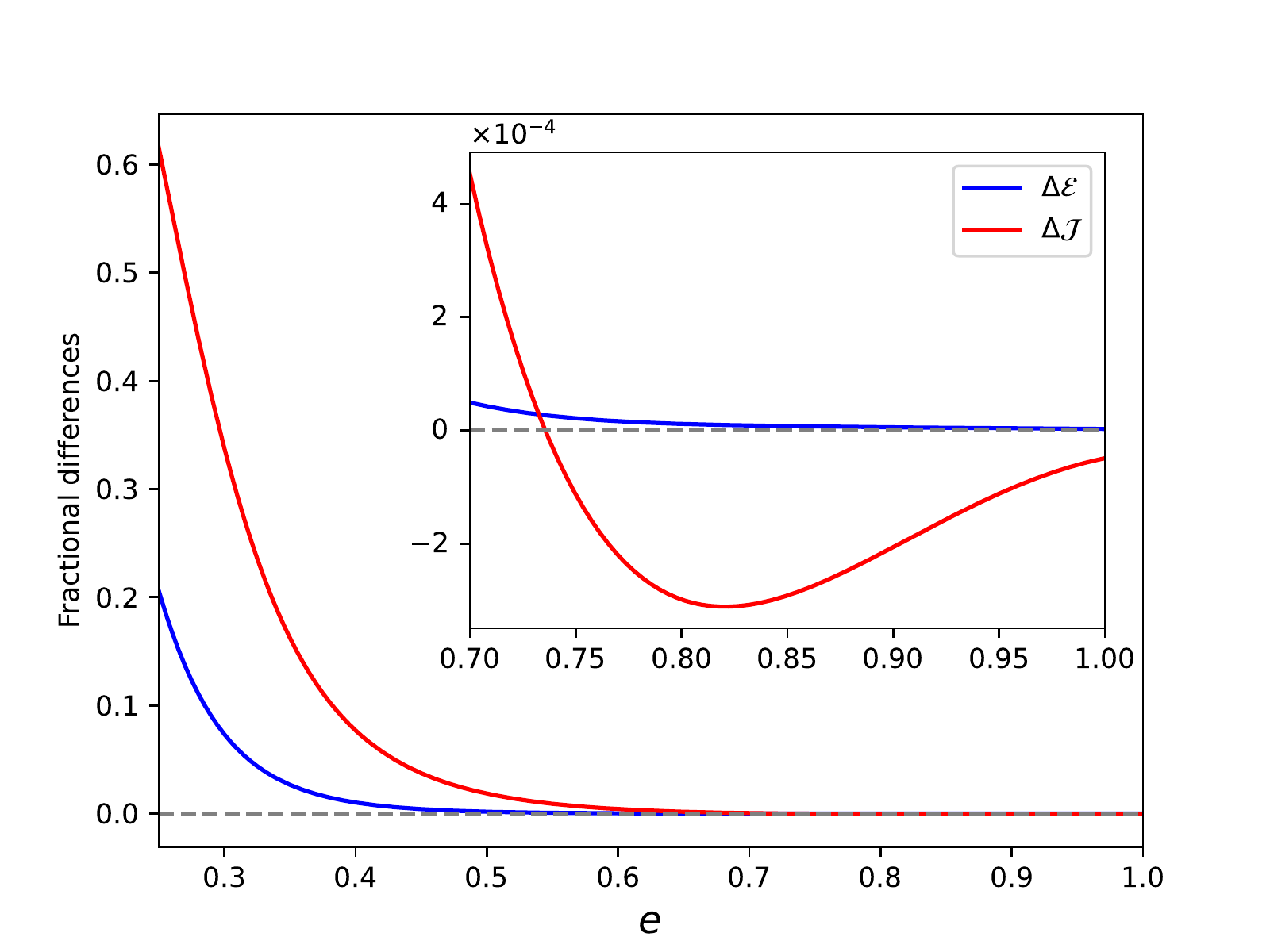}
    \caption{The fractional difference between the 3PN accurate instantaneous fluxes (i.e $\frac{\Delta \mathcal{Q}_\text{hyp}-\Delta \mathcal{Q}_\text{ecc}}{\Delta \mathcal{Q}_\text{hyp}}$, where $\mathcal{Q}= \mathcal{E}, \mathcal{J}$) vs Newtonian eccentricity. Hyperbolic fluxes are expanded in the post-bremsstrahlung limit ($e \to \infty$) while  the elliptic fluxes have been kept in their original forms, namely in terms of $e, \, j, \nu$. The blue curve represents the fractional difference of Energy fluxes while the red curve represents the fractional difference of Angular Momentum fluxes. The zoomed in version focuses on the $0.7$ to $1$ eccentricity range.
     The hyperbolic energy flux, as noted in the text, is expanded up-to $\mathcal{O}(1/{e^5})$ while angular momentum flux has been expanded up-to $\mathcal{O}(1/{e^7})$.
    Additionally, we have chosen $c^2 \, j^2=10$, $\frac{G\,M}{2\,r_0} \, j^2=1/5$ and $\nu=1/4$ for these plots.}
    \label{plot2}
\end{figure}

\begin{figure}[h]
    \centering
    \includegraphics[scale=0.6]{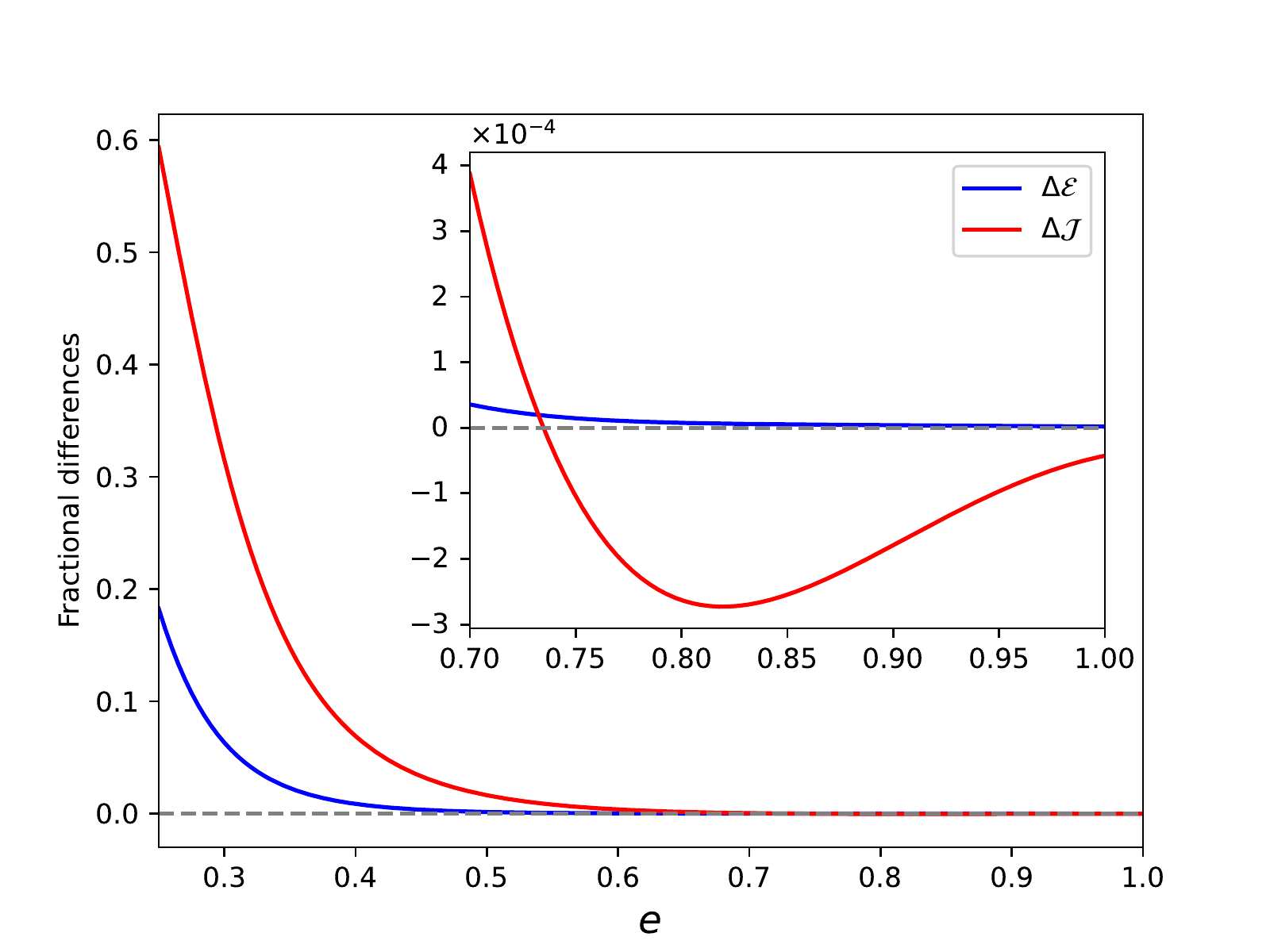}
    \caption{
    Plots that are similar to what are displayed in 
    Fig.~\ref{plot2}. 
    For these plots, we let 
     $c^2 \, j^2=10$, $\frac{G\,M}{2\,r_0} \, j^2=1/5$ and $\nu=1/20$. Clearly, our conclusions are rather independent of $\nu$ values.}
    \label{plot3}
\end{figure}

 In Fig.~\ref{plot2}, we plot the fractional differences between
 Refs.~\cite{arun2008inspiralling, arun2009third} based 
 $\delta {\cal E}$ and $\delta {\cal J}$ 3PN order expression that 
 are exact in $e$
 and the post-bremsstrahlung expansion
 of  our $\Delta \mathcal{E}$, $\Delta \mathcal{J}$ 
 expressions while allowing $e \leq 1$.\\
These plots reveal that the post-bremsstrahlung versions of our  $\Delta \mathcal{E}$, and  $\Delta \mathcal{J}$ expressions provide excellent proxies to compact binaries both in parabolic and high eccentric (bound) orbits.
These  post-bremsstrahlung approximants are substantially different from their eccentric counterparts near their circular 
limits.
This is most likely due to the presence of $\frac{1}{e}$ terms and its multiples and similar conclusions are drawn from plots where
we change values of $\nu$ and the dimensionless 
 $h\,c$ as evident from Fig.~\ref{plot3}.
 We infer that our post-bremsstrahlung $\Delta \mathcal{E}$ and $\Delta \mathcal{J}$ expressions, which should be accurate to describe large eccentric hyperbolic orbit, are not only convergent at the parabolic limit ($e=1$) but also highly eccentric cases ($e\lesssim1$).
 It will be interesting to explore if numerical relativity 
 simulations display a similar behaviour.
This natural convergence at parabolic limit is contrary to the attempt to cover parabolic limit adopted in Sec.~IX in Ref.~\cite{Bini:2017wfr}, where the physical quantities are divergent at parabolic limit, so the parabolic limit is incorporated by numerical methodologies such as fitting and Pade approximation.




\section{Summary and On-going Efforts}\label{summary}

 We have provided explicit expressions for the 3PN-accurate instantaneous contributions to the radiated energy and angular momentum during hyperbolic encounters. These computations, pursued in the time-domain, 
 are not straightforward extensions of the classic 1PN-accurate efforts by Sch\"afer 
 and his collaborators \cite{blanchet1989higher,junker1992binary}.
 This is essentially due to the presence of certain logarithmic terms in the 3PN order contributions to the far-zone flux expressions for the general orbits \cite{arun2008inspiralling}.
 Additionally, we explored the implications of
 post-Bremsstrahlung expansion of our results from the perspective of eccentric orbits.
  \par
There are on-going efforts to compute hereditary contributions to the 
$\Delta {\cal E}$ and $\Delta {\cal J}$ expressions that are accurate to 3PN order, influenced by Ref.~\cite{bini2021gravitational}.
We are also pursuing efforts to compare our PN-accurate results 
with those arising from Numerical Relativity\cite{Damour:2014afa,bae2020gravitational}.
This will be helpful to explore the validity of PN approximation 
while exploring GWs from hyperbolic encounters.
Further, these efforts may allow us to develop a prescription 
for describing GW emission aspects of highly 
eccentric compact binaries with constructs that arise from 
our present and on-going PN-accurate hyperbolic computations.

\section*{Acknowledgments}
We thank Gerhard Sch\"afer and 
Luciano Rezzolla for helpful discussions. G.C is supported by the ERC Consolidator Grant “Precision Gravity: From the LHC to
LISA,” provided by the European Research Council (ERC) under the European Union’s H2020 research and innovation
programme (grant No. 817791).
S.D and A.G acknowledge the support of the Department of Atomic Energy, Government of India, under project identification \# RTI 4002.
A.G is grateful for the financial support and
hospitality of the Pauli Center for Theoretical Studies
and the University of Zurich.
\newpage
\appendix
\begin{widetext}
\section{Hyperbolic Log Integrals}
\label{log_integral}
We provide the details of evaluating certain 3PN order log integrals that are crucial for our results.
We begin with the following expression
\begin{align}
I(x,y):=\int^{+\infty}_{-\infty} \, du \, \frac{\ln(e \cosh u-x)}{e \cosh u -y}.
\end{align}
The goal is to first take a differentiation followd by an integration of $I(x,y)$ with respect to $x$.
Differentiating the above expression with respect to $x$ yields
\begin{subequations}
\begin{align}
-\partial_x \, I(x,y)&=\int^{+\infty}_{-\infty} du \, \frac{1}{e \, \cosh u-x} \, \frac{1}{e \cosh u-y}\,, \nonumber \\[1ex]
&=J_1(x,y)+J_2(x,y),
\end{align}
where
\begin{align}
    J_1(x,y)&:=\frac{2}{\sqrt{e^2-x^2} \, (x-y)} \, \arccos \left( -\frac{x}{e}\right)\,, \\[2ex]
    J_2(x,y)&:=\frac{2}{\sqrt{e^2-y^2} \, (y-x)} \, \arccos \left( -\frac{y}{e}\right)\,.
\end{align}
\end{subequations}
Antiderivative of $-J_2$ with respect to $x$ is 
\begin{align}
I_2:=-\int \, dx\, J_2=\frac{2 \, \ln(y-x) }{\sqrt{e^2-y^2}}\, \arccos \left( -\frac{y}{e}\right)\,.
\end{align}
To integrate  $J_1$, we make a change of variable $f:=\arccos \left( -\frac{x}{e}\right)$, then
\begin{align}
\label{J12}
I_1&:=-\int dx \, J_1=\int df \, \frac{2\,f}{e \cos f+y} =\frac{2\,i}{\sqrt{e^2-y^2}} \, \bigg{[} \text{Li}_2(g)-\text{Li}_2(g^*)\bigg{]}  \nonumber \\[1ex]
& -\frac{2}{\sqrt{e^2-y^2}} \left[\arccos\left(-\frac{y}{e} \right) \, \log \left( - \, \frac{e^2-x \,y+\sqrt{e^2-x^2} \, \sqrt{e^2-y^2}}{4 \, e \, (x-y)}\right)\right]+\alpha(y;e),
\end{align}

where
\begin{subequations}
\begin{align}
g&=\frac{(y-i \, \sqrt{e^2-y^2})}{e} \, \frac{(e+y)\sqrt{e^2-x^2}-(e+x) \sqrt{e^2-y^2}}{(e+y) \, \sqrt{e^2-x^2}+(e+x) \, \sqrt{e^2-y^2}}\,,\\[0.5ex]
g^*&=\frac{(y+i \, \sqrt{e^2-y^2})}{e} \, \frac{(e+y)\sqrt{e^2-x^2}-(e+x) \sqrt{e^2-y^2}}{(e+y) \, \sqrt{e^2-x^2}+(e+x) \, \sqrt{e^2-y^2}}\,.
\end{align}
\end{subequations}

Therein, $\text{Li}_2$ is the polylogarithm function of order 2. These results are valid up to modulo a constant $\alpha(y;e)$ that is independent of $x$. To determine the expression of $\alpha$, let us consider $x \to - \infty$ behavior such that
\begin{align}
    \lim_{x \to - \infty} I(x,y)=\lim_{x\to-\infty}\frac{2 \, \log(-x)}{\sqrt{e^2-y^2}} \, \arccos \left( -\frac{y}{e}\right).
\end{align}
An, it can be easily checked that
\begin{align}
\lim_{x \to - \infty} I(x,y)=\lim_{x \to - \infty} I_2(x,y)\,,
\end{align}
which implies
\begin{align}
\lim_{x \to - \infty} I_1(x,y)=0\,,
\end{align}
to keep $I=I_1+I_2$ hold.
Making use of this condition at $x\to-\infty$ on Eq. \eqref{J12}, we reach
\begin{align}
\alpha(y;e)=\frac{2 \, i}{\sqrt{e^2-y^2}} \left\{ \frac{\pi^2}{6}-\text{Li}_2 \left[ -\frac{(i \, y+\sqrt{e^2-y^2}\, )^2}{e^2}\right]-i \, \arccos \left( -\frac{y}{e}\right) \, \log \left[ \frac{4 \, (y- i\, \sqrt{e^2-y^2})}{e}\right] \right\}.
\end{align} 
In the above we used $\lim_{x\to-\infty}\text{Li}_2(g^*)=\frac{\pi^2}{6}$.
Thus, we finally obtain
\begin{align}
\allowdisplaybreaks
\label{fl}
    I&=\frac{2 \, i}{\sqrt{e^2-y^2}} \bigg{\{}\frac{\pi^2}{6}+ \text{Li}_2(g)-\text{Li}_2(g^*)-\text{Li}_2\left(-\frac{(i y+\sqrt{e^2-y^2})^2}{e^2}\right)-i \, \arccos \left(-\frac{y}{e}\right) \notag \\[1ex]
    & \times \log \Bigg{[} \frac{(y-i\, \sqrt{e^2-y^2}) \,(e^2-x \, y+\sqrt{e^2-x^2} \, \sqrt{e^2-y^2}) }{e^2}\Bigg{]} \Bigg{\}}.
\end{align}
Now, we focus on a special condition on Eq. \eqref{fl} where $x \to y$, the integral we are interested to compute.
Note that when $y \to x$, there are some divergences because of $\sim \, \log(y-x)$ term in $I_1$, but with proper calculus, these divergences 
are cancelled while taking $y \to x$ limit.
 we get
\begin{align}
\allowdisplaybreaks
\label{fla}
\int_{-\infty}^\infty \, du \, \frac{\ln(e \cosh u-x)}{e \cosh u -x}&=\frac{2 \, i}{\sqrt{e^2-x^2}} \, \Bigg{\{} \frac{\pi^2}{6} -\text{Li}_2 \left( -\frac{(i \, x+\sqrt{e^2-x^2})^2}{e^2} \right)-i\, \arccos\left(-\frac{x}{e}\right) \notag \\[1ex]
&\times \, \log \left[ \frac{2 \, (e^2-x^2) \, (x-i\, \sqrt{e^2-x^2})}{e^2}\right]\Bigg{\}}.
\end{align}

Eq. \eqref{fla} can be further simplified using the usual definition of Clausen function  $\text{Cl}_2(x)$ \cite{lewin1991structural} of order 2 -  given by the integral
\begin{align}
\text{Cl}_2(x)=- \int_0^x dx \,  \log \left|2 \sin \frac{x}{2}\right|
\end{align}
There exist a nice expression of Dilogaritm function :$\text{Li}_2(x)$ in terms of  Clausen function, given by
\begin{align}
\label{ep}
\text{Li}_2(\text{e}^{i \, \theta})=\frac{\pi^2}{6}-\frac{1}{4} \, |\theta| (2 \pi-|\theta|)+i \,\text{Cl}_2(\theta), \enspace |\theta|\leq 2 \pi.
\end{align}
By virtue of Eq. \eqref{ep} we  obtain the most simplified form of the integral $I(x)$. 
Our final expression is 
\begin{align}
\int_{-\infty}^\infty \, du \, \frac{\ln(e \cosh u-x)}{e \cosh u -x}=\frac{2}{\sqrt{e^2-x^2}} \left\{\text{Cl}_2 \left(-\frac{x}{e}\right)+\arccos \left(-\frac{x}{e}\right) \ln \left(2 \,  e-\frac{2 \, x^2}{e}\right)\right\}.
\end{align}
\\

\section{
Energy and angular momentum radiations in terms of energy and angular momentum
} 
\label{appendix_AC2}
Since the time eccentricity $e_t$ is not gauge invariant, we provide maximally gauge invariant expressions of $\Delta \mathcal{E}$ and $\Delta\mathcal{J}$ in terms of energy $E$ and angular momentum $j$. Note that it could be never possible to provide fully gauge invariant expression, because instantaneous part of $\Delta \mathcal{E}$, $\Delta \mathcal{J}$ is not gauge invariant, since they depend on the separation of the scale $r_0$ differentiating short/long scales. Except the issue of $r_0$ which will be removed by hereditary contribution, the following expressions are gauge invariant. Here, we use the Newtonian eccentricity $e$ as a shorthand symbol of $\sqrt{1+2\,E\,j^2}$ without implying any geometric meaning.
 

\begin{align}
\Delta \mathcal{E}&=\frac{2}{15} \,\frac{M \, \nu ^2}{j^7 \, c^5} \Big{(} \mathcal{I}_N^\text{MH}+\frac{1}{j^2 \, c^2} \,  \mathcal{I}_\text{1PN}^\text{MH}+\frac{1}{j^4 \, c^4} \,  \mathcal{I}_\text{2PN}^\text{MH}+\frac{1}{j^6 \, c^6} \,  \mathcal{I}_\text{3PN}^\text{MH} \Big{)}\,,
\end{align}
where 
\begin{subequations}
\label{Ie}
\begin{align}
\mathcal{I}_N^\text{MH}&=\sqrt{e^2-1} \, \Bigg{[}\frac{602}{3}+\frac{673 e^2}{3}\Bigg{]}+\arccos \left(-\frac{1}{e} \right) \, \Bigg{[}96+292 \, e^2+37 \,  e^4\Bigg{]} \,,\\[2ex]
\mathcal{I}_\text{1PN}^\text{MH}&=\sqrt{e^2-1} \Bigg{[}\frac{108+12 \, \nu
}{e^2}+\Bigg{(}\frac{90754}{35}-\frac{2185 \, \nu }{6}\Bigg{)}+e^2 \, 
\Bigg{(}\frac{141439}{70}-\frac{11441 \, \nu }{6}\Bigg{)}+e^4 \, \Bigg{(}\frac{89907}{280}-\frac{1117 \, \nu }{2}\Bigg{)}\Bigg{]} \notag \\[1ex]
&+\arccos \left(-\frac{1}{e} \right) \Bigg{[}\Bigg{(}\frac{11177}{7}-95 \, \nu \Bigg{)}+e^2 \,  \Bigg{(}\frac{37785}{14}-\frac{3051
	\, \nu }{2}\Bigg{)}+e^4 \, \Bigg{(}\frac{2817}{4}-\frac{2283 \, \nu }{2}\Bigg{)}+e^6 \,  \Bigg{(}\frac{2393}{56}-\frac{111 \, \nu }{2}\Bigg{)} \Bigg{]} \,,\\[2ex]
\mathcal{I}_\text{2PN}^\text{MH}&=\sqrt{e^2-1} \Bigg{[}\frac{1}{e^4} \, \Bigg{(}-\frac{243}{2}-27 \, \nu 
		-\frac{3 \, \nu ^2}{2}\Bigg{)}+\frac{1}{e^2}\, \Bigg{(}\frac{207423}{112}-\frac{2515
			\, \nu }{56}-\frac{179 \, \nu ^2}{16} \Bigg{)}+\Bigg{(}\frac{8409586747}{317520}\notag\\[1ex]
&-\frac{9858601 \, \nu }{720}-\frac{1569 \, \nu ^2}{4}\Bigg{)}+e^2 \,  \Bigg{(}\frac{10125905183}{635040}-\frac{38505773 \, \nu }{1440}+4098 \, \nu ^2\Bigg{)}+e^4 \,  \Bigg{(}-\frac{154205167}{158760}-\frac{36653233
		\, \nu }{5040}\notag\\[1ex]
&+\frac{11581 \, \nu ^2}{2}\Bigg{)}+e^6  \, \Bigg{(}\frac{509759}{1960}-\frac{680471 \, \nu }{1120}+\frac{12693
		\, \nu ^2}{16}\Bigg{)}\Bigg{]}+\arccos \left(-\frac{1}{e} \right) \Bigg{[} \frac{52925837}{3024}-\frac{158383 \, \nu }{24}-\frac{4787 \, \nu ^2}{16}\notag\\[1ex]
&+e^2 \,  \Bigg{(}\frac{36730439}{1512}-\frac{4415855 \, \nu }{168}+\frac{3877 \, \nu ^2}{2}\Bigg{)}+e^4 \,  \Bigg{(}\frac{479005}{288}-\frac{3069361 \, \nu }{224}+\frac{25641 \, \nu ^2}{4}\Bigg{)}+e^6 \,  \Bigg{(}-\frac{5885}{84}\notag\\[1ex]
&-\frac{3489
\, \nu }{2}+\frac{4339 \, \nu ^2}{2}\Bigg{)}+e^8 \, \Bigg{(}\frac{745}{12}-\frac{11965\, \nu }{224}+\frac{925 \, \nu ^2}{16}\Bigg{)}\Bigg{]} \,,\\[2ex]
\mathcal{I}_\text{3PN}^\text{MH}&=\sqrt{e^2-1}\Bigg{[}\frac{6713608}{1575}+\frac{17868572 \, 
		e^2}{525}+\frac{19300553 \, e^4}{525} +\frac{17525209 e^6}{3150}\Bigg{]}\, \log \left[ \frac{2\,r_0\,(e^2-1)}{G\,M \, j^2 \, e} \right]-\Bigg{[}\frac{54784}{35} \notag \\[1ex]
&+\frac{465664 \, e^2}{21}+\frac{4426376 \, e^4}{105}
	+\frac{1498856 \, e^6}{105}+\frac{31779 \, e^8}{70}\Bigg{]} \, 
\Bigg{\{} \log \left[ \frac{2 \, G\,M \, j^2}{r_0\,e} \right] \, \arccos \left(-\frac{1}{e} \right) \notag\\[1ex]
&+\text{Cl}_2 \left[ 2\, \arccos \left(-\frac{1}{e} \right) \right]\Bigg{\}}
+\sqrt{e^2-1} \, \Bigg{[}\frac{1}{4 \, e^6} \, \Bigg{(}729 +243 \, \nu+27 \, \nu^2+\, \nu^3\Bigg{)} +\frac{1}{e^4} \, \Bigg{(}-\frac{1106055}{448}\notag\\[1ex]
&-\frac{63027 \, \nu }{448}+\frac{10159 \, \nu ^2}{448}+\frac{167 \, \nu ^3}{192}\Bigg{)}+\frac{1}{e^2} \, \Bigg{(}\frac{15311843}{672}+\Bigg{(}-\frac{2862985}{378}+\frac{123\pi ^2}{2}\Bigg{)} \, \nu 	-\frac{426791 \, \nu ^2}{672}-\frac{79 \, \nu ^3}{4}\Bigg{)}\notag\\[1ex]
&+\frac{40066564486859}{139708800}+\Bigg{(}-\frac{205531241131}{635040}	+\frac{13414913
		\pi ^2}{3360}\Bigg{)} \, \nu +\frac{833407139 \, \nu ^2}{40320}+\frac{206305 \, \nu ^3}{192}\notag\\[1ex]
&+e^2 \, \Bigg{(}\frac{8653194752669}{23284800}+\Bigg{(}-\frac{243329285581}{423360}+\frac{2172959
		\pi ^2}{280}\Bigg{)} \, \nu +\frac{310440959 \, \nu ^2}{2240}-\frac{4703 \, \nu ^3}{48}\Bigg{)}\notag\\[1ex]
&+e^4 \, \Bigg{(}\frac{10874967898483}{46569600}+\Bigg{(}-\frac{13311394987}{141120}-\frac{995521
		\pi ^2}{896}\Bigg{)} \, \nu 	+\frac{131367287 \, \nu ^2}{1344}-\frac{1086249 \, \nu ^3}{64}\Bigg{)}\notag\\[1ex]
&+e^6 \, \Bigg{(}\frac{723666251779}{19958400}+\Bigg{(}\frac{19905131933}{1270080}-\frac{12303239
		\pi ^2}{13440}\Bigg{)} \, \nu +\frac{284599649 \, \nu ^2}{20160}	-\frac{261641 \, \nu ^3}{24}\Bigg{)}+e^8 \Bigg{(}\frac{248312131}{4139520}\notag\\[1ex]
&-\frac{11427029 \, \nu
	}{28224}+\frac{3374859 \, \nu ^2}{4480}-\frac{56047 \, \nu ^3}{64}\Bigg{)} \Bigg{]}+\arccos \left(-\frac{1}{e} \right) \Bigg{[}\frac{642008180503}{3326400}+\Bigg{(}-\frac{21604997}{126}+\frac{15785
		\pi ^2}{8}\Bigg{)} \, \nu \notag\\[1ex]
&+\frac{7846337 \, \nu ^2}{1344}+\frac{12633 \, \nu ^3}{32}	+e^2 \, \Bigg{(}\frac{2827833857903}{6652800}	+\Bigg{(}-\frac{6912058987}{12096}+\frac{124763 \pi ^2}{16}\Bigg{)}
	\, \nu +\frac{266844577 \, \nu ^2}{2688}\notag\\[1ex]
&+\frac{26815 \, \nu ^3}{16}\Bigg{)}+e^4 \, \Bigg{(}\frac{1246592422921}{3326400}	+\Bigg{(}-\frac{3120760105}{12096}+\frac{112709
		\pi ^2}{64}\Bigg{)} \, \nu +\frac{174794687 \, \nu ^2}{1344}-\frac{91945 \, \nu ^3}{8}\Bigg{)}\notag\\[1ex]
&+e^6 \, \Bigg{(}\frac{31441963853}{277200}+\Bigg{(}\frac{86715535}{6048}-\frac{211027\, \pi ^2}{128}\Bigg{)} \, \nu +\frac{7483281 \, \nu ^2}{224}-\frac{245475 \, \nu ^3}{16}\Bigg{)}+e^8 \,  \Bigg{(}\frac{5518709201}{1478400}+\Bigg{(}\frac{2020985}{1344}	\notag\\[1ex]
&-\frac{12177
		\pi ^2}{128}\Bigg{)} \, \nu +\frac{296053 \, \nu ^2}{112}-\frac{95255 \, \nu ^3}{32}\Bigg{)}+e^{10} \,  \Bigg{(}\frac{2075735}{118272}-\frac{745 \, \nu }{12}+\frac{45467
		\, \nu ^2}{896}-\frac{407 \, \nu ^3}{8}\Bigg{)}\Bigg{]}.
\end{align}
\end{subequations}
Similarly, the 3PN-accurate 
instantaneous contributions to the radiated 
angular momentum  terms of  $e$ and $j$ become 
\begin{align}
\Delta \mathcal{J}&=\frac{8}{5 \, c^5} \, \frac{G M^2\, \nu^2}{ j^4}  \Big{(} \mathcal{H}_N^\text{MH}+\frac{1}{j^2 \, c^2} \,  \mathcal{H}_\text{1PN}^\text{MH}+\frac{1}{j^4 \, c^4} \,  \mathcal{H}_\text{2PN}^\text{MH}+\frac{1}{j^6 \, c^6} \,  \mathcal{H}_\text{3PN}^\text{MH}\Big{)}\,,
\end{align}
where 
\begin{subequations}
\label{He}
\begin{align}
\mathcal{H}^\text{MH}_\text{N}&=\sqrt{e^2-1}\,\Bigg{[}13+2 e^2\Bigg{]}+\arccos \left( -\frac{1}{e}\right) \, \Bigg{[}8+7 e^2 \Bigg{]}\,,\\[2ex]
\mathcal{H}^\text{MH}_\text{1PN}&= \sqrt{e^2-1} \, \Bigg{[}\frac{9+\, \nu }{e^2}+\Big{(}\frac{11525}{168}-\frac{1451 \, \nu }{36}\Big{)}+e^2 \, \Bigg{(}\frac{10313}{336}-\frac{2065 \, \nu
}{36}\Bigg{)}+e^4 \, \Bigg{(}\frac{109}{28}-\frac{5 \, \nu }{4}\Bigg{)}\Bigg{]}\notag\\[1ex]
&+\arccos \left( -\frac{1}{e}\right) \Bigg{[}\frac{4877}{84}-\frac{65 \, \nu }{4}+e^2 \, \Bigg{(}\frac{869}{21}-\frac{419 \, \nu }{6}\Bigg{)}+e^4 \,
\Bigg{(}\frac{4283}{336}-\frac{71 \, \nu }{6}\Bigg{)}\Bigg{]}  
\,,\\[2ex]
\mathcal{H}^\text{MH}_\text{2PN} &= \sqrt{e^2-1} \Bigg{[}-\frac{1}{8 \, e^4} \Bigg{(}81+ 18\, \nu+\, \nu^2 \Bigg{)}+\frac{1}{e^2} \,  \Bigg{(}\frac{41169}{448}-\frac{11839
		\, \nu }{672}-\frac{109 \, \nu ^2}{64}\Bigg{)}+\frac{139894943}{272160}-\frac{17944223 \, \nu }{30240}\notag\\[1ex]
&+\frac{1915 \, \nu ^2}{144}+e^2 \,  \Bigg{(}-\frac{18473111}{272160}-\frac{201089
	\, \nu }{540}+\frac{25595 \, \nu ^2}{96}\Bigg{)}+e^4 \, \Bigg{(}-\frac{13557}{560} 
-\frac{870131 \, \nu }{10080}+\frac{15701 \, \nu ^2}{144}\Bigg{)}\notag\\[1ex]
&+e^6 \, \Bigg{(}\frac{3245}{4032}-\frac{327 \, \nu
}{224}+\frac{51 \, \nu ^2}{64}\Bigg{)} \Bigg{]}+
\arccos \left( -\frac{1}{e}\right)
\Bigg{[}\frac{2051491}{4536}-\frac{46297 \, \nu }{126}-\frac{79 \, \nu ^2}{12}+e^2
\Bigg{(}\frac{752767}{6048}\notag\\[1ex]
&-\frac{121675 \, \nu }{224}
+\frac{4433 \, \nu ^2}{24}\Bigg{)}+e^4 \,  \Bigg{(}-\frac{28297}{378}-\frac{24611 \, \nu }{168}+\frac{783
	\, \nu ^2}{4}\Bigg{)}+e^6 \, \Bigg{(}\frac{1327}{504}-\frac{3653 \, \nu }{224}+\frac{337 \, \nu ^2}{24}\Bigg{)}\Bigg{]}\,,\\[2ex]
\mathcal{H}^\text{MH}_\text{3PN}&=\sqrt{e^2-1} \, \Bigg{[}\frac{99724}{315}+\frac{351067 e^2}{315}+\frac{210683 e^4}{630}\Bigg{]} \, \log \left[ \frac{r_0\,(e^2-1)}{G\,M \, j^2 \, e} \right]-\Bigg{[}\frac{13696}{105}+\frac{98012 e^2}{105}+\frac{23326 e^4}{35}+\frac{2461
	e^6}{70}\Bigg{]} \notag\\[1ex]
& \times \Bigg{\{} \log \left[ \frac{2 \, G\,M \, j^2}{r_0\,e} \right] \, \arccos \left(-\frac{1}{e} \right)+\text{Cl}_2 \left[ 2\, \arccos \left(-\frac{1}{e} \right) \right]\Bigg{\}}+\sqrt{e^2-1} \, \Bigg{[} \frac{1}{16 \, e^6} \Bigg{(} 243+81 \, \nu+ 9 \, \nu^2+\, \nu^3\Bigg{)}\notag\\[1ex]
&+\frac{1}{e^4} \Bigg{(}-\frac{257553}{1792} 
+\frac{16035\, \nu }{1792}+\frac{22507 \, \nu ^2}{5376}+\frac{121 \, \nu ^3}{768} \Bigg{)}+\frac{1}{e^2} \, \Bigg{(}\frac{27158603}{32256}
+\Bigg{(}-\frac{165077725}{290304}+\frac{41 \pi ^2}{8}\Bigg{)} \, \nu \notag\\[1ex]
&-\frac{718787 \, \nu ^2}{32256}-\frac{237 \, \nu ^3}{512}\Bigg{)}+\frac{387361378703}{67060224}+\Bigg{(}-\frac{337799063429}{30481920}+\frac{340423 \pi
	^2}{1920}\Bigg{)} \, \nu +\frac{878650477 \, \nu ^2}{483840}+\frac{344695 \, \nu ^3}{4608}\notag\\[1ex]
&+e^2 \Bigg{(}\frac{1005116464409}{167650560}+\Bigg{(}-\frac{23645568707}{3048192}+\frac{512213
	\pi ^2}{3840}\Bigg{)} \, \nu +\frac{134389529 \, \nu ^2}{34560}-\frac{708305 \, \nu ^3}{2304}\Bigg{)}+e^4 \Bigg{(}\frac{432742948027}{335301120}\notag\\[1ex]
&+\Bigg{(}\frac{5045091193}{7620480}-\frac{44567
	\pi ^2}{1920}\Bigg{)} \, \nu +\frac{60548279 \, \nu ^2}{60480}-\frac{886135 \, \nu ^3}{1152}\Bigg{)}+e^6 \Bigg{(}-\frac{1172569271}{24837120}
+\frac{17091961 \, \nu }{1128960}+\frac{4248733 \, \nu ^2}{32256}\notag\\[1ex]
&-\frac{687113 \, \nu ^3}{4608}\Bigg{)} +e^8 \Bigg{(}-\frac{6973}{354816}-\frac{3245 \, \nu }{32256}+\frac{3379 \, \nu ^2}{3584}-\frac{249 \, \nu ^3}{512}\Bigg{)} \Bigg{]}+\arccos \left( -\frac{1}{e}\right) \Bigg{[}\frac{26951805241}{5702400}\notag\\[1ex]&
+\Bigg{(}-\frac{503311993}{72576}+\frac{3239
	\pi ^2}{32}\Bigg{)} \, \nu +\frac{12801961 \, \nu ^2}{16128}+\frac{2479 \, \nu ^3}{64}+e^2\Bigg{(}\frac{83623027931}{9979200}+\Bigg{(}-\frac{408126835}{36288}+\frac{6355\pi ^2}{32}\Bigg{)} \, \nu \notag\\[1ex]
	&+\frac{15335105 \, \nu ^2}{4032}-\frac{3907 \, \nu ^3}{48}\Bigg{)}+e^4 \Bigg{(}\frac{6781095059}{1663200}
+\Bigg{(}-\frac{18560153}{24192}-\frac{615
	\pi ^2}{256}\Bigg{)} \, \nu +\frac{5148791 \, \nu ^2}{2688}-\frac{71389 \, \nu ^3}{96}\Bigg{)}\notag\\[1ex]
	&+e^6 \Bigg{(}\frac{254451209}{3326400}+\Bigg{(}\frac{1448255}{6048}-\frac{615
	\pi ^2}{128}\Bigg{)} \, \nu 
+\frac{391981 \, \nu ^2}{1344}-\frac{16867 \, \nu ^3}{48}\Bigg{)}
+e^8 \Bigg{(}-\frac{227005}{709632}-\frac{14411 \, \nu }{2016}+\frac{29727
	\, \nu ^2}{1792}\notag\\[1ex]
	&-\frac{2693 \, \nu ^3}{192}\Bigg{)}\Bigg{]}.
	\end{align}
\end{subequations}
\end{widetext}
\nocite{*}
\bibliography{mybib}

\end{document}